\documentclass[aip, reprint]{revtex4-1}
\usepackage[dvips]{graphicx}
\usepackage{color}
\usepackage{amssymb,amsmath,amsfonts}

\begin{document}

\title{Reynolds Number Effects on Mixing Due to Topological Chaos}

\author{Spencer A. Smith}
\author{Sangeeta Warrier}
\affiliation{	Department of Physics, Mount Holyoke College, South Hadley, MA 01075, USA}

\begin{abstract}

\large
Topological chaos has emerged as a powerful tool to investigate fluid mixing.  While this theory can guarantee a lower bound on the stretching rate of certain material lines, it does not indicate what fraction of the fluid actually participates in this minimally mandated mixing.  Indeed, the area in which effective mixing takes place depends on physical parameters such as the Reynolds number.  To help clarify this dependency, we numerically simulate the effects of a batch stirring device on a 2D incompressible Newtonian fluid in the laminar regime.  In particular, we calculate the finite time Lyapunov exponent (FTLE) field for three different stirring protocols, one topologically complex (pseudo-Anosov) and two simple (finite-order), over a range of viscosities.  After extracting appropriate measures indicative of both the amount of mixing and the area of effective mixing from the FTLE field, we see a clearly defined Reynolds number range in which the relative efficacy of the pseudo-Anosov protocol over the finite-order protocols justifies the application of topological chaos.  More unexpectedly, we see that while the measures of effective mixing area increase with increasing Reynolds number for the finite-order protocols, they actually exhibit non-monotonic behavior for the pseudo-Anosov protocol.
 
\end{abstract}

\maketitle

\begin{quotation}

The intuitive notion of fluid mixing can be quantified by global measures such as the topological entropy of flow maps, as well as local measures like finite-time Lyapunov exponents.  Each can be calculated if given sufficiently detailed knowledge of the advected fluid-particle trajectories, however, experimental or computational limitations might preclude this.  Fortunately, if the fluid system has braid-like topological obstructions, such as stirring rods, then we can use the ideas of topological chaos to provide a lower bound on the stretching rate of certain material lines.  This, in turn, forces a lower bound on our mixing measure, the finite time Lyapunov exponent, on some fraction of the fluid. This minimally mandated mixing is computable using the algebraic information associated with each braid, and is therefore very useful when confronted with limited fluid flow data.  This great simplification comes at a price - the actual mixing could be much larger than the lower bound.  Additionally, the area where the fluid is forced to have at least this minimally mandated mixing is not determined by the topological theory, and could be vanishingly small.  Ultimately, the degree to which these two issues affect the usefulness of applying topological ideas to fluid mixing will depend on physical parameters of the system, such as the Reynolds number. 

\end{quotation}

\section{Initial Discussion}

Fluid mixing is one of the rare arenas where chaotic motion is not an unwanted phenomenon.  Indeed, chaotic advection can help to quantify and optimize the mixing in many industrial applications\cite{FinnCoxMix} where the goal is to efficiently combine two or more substances.  Similar ideas can also enhance the mixing of polymers, biological samples, or chemicals in microfluidic devices\cite{2014arXiv1403.2953A}, where low Reynolds numbers preclude turbulent mixing.  Even on the much larger scale of oceanic flows, chaotic advection can help to measure the mixing of nutrient concentration, salinity, and temperature\cite{Samelson13,Thiffeault:2010ir}.  These diverse examples demonstrate the broad application of chaotic advection, which was introduced by Aref\cite{Aref:1984} over three decades ago.

However, in many systems, the presence of chaotic advection can be reduced or even outright removed by a simple change of physical parameters \cite{bib:Aref1}, and is therefore not a fully robust attribute.  Fortunately, there exists a large class of systems in which topological barriers force a certain minimal amount of complexity to be present in the fluid flow, regardless of the value of physical parameters.  This robust, topologically-induced, chaos allows us to estimate the minimal mixing in the absence of detailed knowledge of the actual fluid flow.  Simple algebraic characteristics of the topological obstructions are sufficient to calculate the lower bound on the mixing forced upon portions of the fluid. 

The study of topological chaos, starting with Boyland\cite{Boyland:2000uc,Boyland:1994ud}, has been very fruitful in the past decade and a half\cite{Thiffeault:2006jp,Thiffeault:2005hn,Thiffeault:2008er,Finn200692}.  It has branched out from systems with obvious topological obstructions, such as rods in a batch stirring device\cite{Finn:2010wg}, to more general systems through the recognition that periodic orbits, acting as ``ghost rods"\cite{Gouillart:2006hi}, can serve as topological barriers.  Even the restriction that the topological obstructions be part of a strictly periodic motion has been loosened to include quasi-periodic motion\cite{Stremler:2011hu}.  Indeed, the generically non-periodic motion of collections of arbitrary advected fluid-particles can be used to estimate the topological entropy of a natural flow\cite{Thiffeault:2010ir}.  As we will show, topological chaos is most useful in the low Reynolds number regime of Stokes flow, however it has also been used to great effect in classifying the inviscid flow of collections of point vortices\cite{Boyland:2003ux}.  Aside from its many applications, topological chaos is a beautiful mathematical theory that touches upon varied disciplines such as dynamic systems theory, algebraic topology, and fluid dynamics.

At the heart of topological chaos are two issues that are problematic for applying the theory to real fluids.  First, the actual amount of mixing in the fluid could be much larger than the topological lower bound might suggest.  In the Stokes limit of low Reynolds number, the motion of the fluid and that of the topological obstructions are well entrained, and it is natural to expect that the lower bound is close to the actual mixing.  However, as the Reynolds number is increased, the inertial motion of the fluid results in excess mixing.  To account for this, one can include successively more periodic orbits in the set of topological obstructions and the calculated lower bound on mixing will approach that of the actual fluid\cite{Gouillart:2006hi,Stremler2012}.  This requires sufficient knowledge of the fluid motion to allow the calculation of periodic orbits.  If only the motion of the original topological obstructions is known, then it would be useful to know the range in Reynolds number over which the topological theory gives a useful lower bound.  The second issue involves the fluid area in which the mixing is at least that of the lower bound.  The topological theory does not quantify this area, and indeed the mixing lower bound could be enforced on a vanishingly small fraction of the total fluid domain.  Certainly the topologically induced lower bound is useless if it does not reflect the mixing in the majority of the fluid.  Both Vikhansky\cite{Vikhansky:2004} and Finn, Cox, and Byrne\cite{Finn:2003hd} have investigated this issue, though only in the Stokes or inviscid regimes and without a quantitative measure of the area participating in a high level of mixing.

To investigate both of these issues and their dependance on Reynolds number, we numerically simulated (see \ref{ssec:NumExp}) laminar fluid flow due to a batch stirring device (see \ref{ssec:BSD}).  We varied the stirring protocol to represent three different braids (see \ref{ssec:TopTh}), one topologically complex and two topologically simple (see fig.~(\ref{graph:Braidpic})).  Likewise, we varied the viscosity of the fluid to give a range of Reynolds numbers.  For each simulation, we calculated the finite-time Lyapunov exponent (FTLE) field and extracted measures indicative of the maximum mixing and the area of efficient mixing (see \ref{ssec::MixMeasure}).  

\section{Background}
	
	\subsection{Batch Stirring Device} \label{ssec:BSD}
	
\begin{figure}[htbp]
 \centering
  \large
   \includegraphics{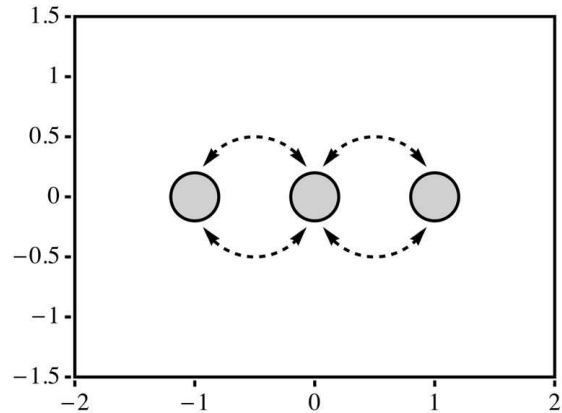}
  \normalsize
    \caption{Batch stirring device geometry and pairwise circular motion of the rods.}
    \label{graph:Schematic}
\end{figure}

Batch stirring devices consist of a fluid domain and a set of parallel stirring rods which execute periodic motion.  When there is an insignificant amount of fluid transport in the vertical direction, we can consider the fluid domain to be effectively two dimensional.  Due to its simplicity and its ability to create complex topology through the braided motion of the stirring rods, this system is well established as an arena to test out new ideas\cite{Boyland:2000uc}.  More specifically, we consider a 2D incompressible Newtonian fluid interacting with three stirring rods in a $4\times3$ rectangular domain.  The stirring rods are disks of radius $0.2$, with centers lying on the long axis and separated by a distance of $1$.  Note that we do not specify the units of length, since the absolute size of the system is not important.  The physics of the flow is dictated by the Reynolds number, the ratio of a characteristic distance and velocity to the kinematic viscosity.  See fig.~(\ref{graph:Schematic}) for the geometry of this batch stirring device.

The overall motion of the stirring rods is comprised of a succession of neighboring pairs of rods switching positions.  For each switch the pertinent pair of rods execute circular motion about their relative center, either in a clockwise (CW) or counter-clockwise (CCW) direction.  A stirring protocol with a given number of switches is then specified by the location of each switch (left pair or right pair), the type of each switch (CW or CCW), and the order in which they occur.  We will naturally identify stirring protocols with an element of the braid group (see \ref{ssec:TopTh}).  During a switch, each rod moves with an average speed of $0.5\;units/s$.  However, the rods do not move at a constant speed, as this would create large accelerations that are not physically possible when a rod starts or stops moving.  Instead, the distance a rod has moved along its path during a switch is parameterized by a rescaled time, $t \rightarrow T_g \left(1-\cos\left(\pi t/T_g\right)\right)/2$, which preserves average speed and leads to physically realistic movement.  Here, $T_g$ is the time to execute one switch, which in our case is $\pi$ seconds.	
			
	\subsection{Topological Theory} \label{ssec:TopTh}
	
\begin{figure}[htbp]
 \centering
  \large
   \includegraphics{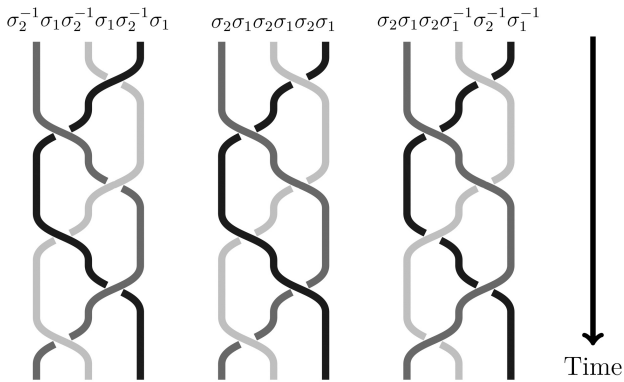}
  \normalsize
    \caption{Braids from left to right: pseudo-Anosov ``golden" braid, finite-order (full twist), and finite-order (isotopic to the identity).}
    \label{graph:Braidpic}
\end{figure}

The topological structures relevant to the batch stirring device are geometric braids.  Consider the direct product of the 2D fluid domain with a time dimension flowing downward.  In this domain, the periodic motion of the stirring rods trace out the strands of a braid.  Once we fix the geometric movement of each rod as it switches position (see \ref{ssec:BSD}), all that is left to completely characterize the stirring protocol is the braid topology.  We can compactly specify our stirring protocol by describing each braid with its Artin braid generators.  Switching rods $i$ and $i+1$, labeled left to right, in a counter-clockwise manner is given by the positive braid generator $\sigma_i$, while switching them clockwise is given by $\sigma_i^{-1}$, the inverse of $\sigma_i$.  Now a stirring protocol can be specified by a string of generators, read left to right, which constitute a braid word (e.g. see fig.~(\ref{graph:Braidpic})).  Briefly, we must note that there are many conventions for describing braids.   We have followed Birman \cite{bib:BirmanBLMCG}, whereas others \cite{Boyland:2000uc,Thiffeault:2006jp} instead take time to flow upwards, which is equivalent to switching the association of the positive generator from CCW to CW motion and similarly for its inverse.  
	
For a given stirring protocol, the associated simple algebraic braid word carries a lot of information.  Among other things, the braid will determine how much mixing is forced on the surrounding fluid by the movement of the stirring rods.  To make this idea more concrete, first consider the flow map, $f^T\left( \mathbf{x}_0 \right) = \mathbf{x}\left( T\right)$, which takes points in the fluid domain and maps them to their final position after being passively transported by the fluid flow for a time $T$.  It is the topological entropy, $h\left( f^T \right)$, or finite-time Lyapunov exponent field, $\Lambda\left(\mathbf{x}_0, T\right)$, of this map which quantifies the amount of mixing that occurs in this fluid.  Including the flow map, there are many maps that are consistent with the given braid word topology, and therefore mutually isotopic.  We can group all of these maps together and refer to them collectively as a mapping class.  The space of all maps on our domain is partitioned into disjoint mapping classes, each one associated with a different braid.  Indeed, the mapping class group\cite{bib:BirmanBLMCG} of our $3$-punctured fluid domain is equivalent to the braid group, $B_3$, on three strands.

In each mapping class there is a particular map, the minimal representative, $f^{*}$, which has the minimum amount of complexity.  The complexity is measured by the topological entropy, which tracks the exponential growth rate of distinguishable orbits as time increases.  Interestingly, a whole class of braids, pseudo-Anosov (pA) braids, have minimal maps with non-zero topological entropy.  Through Handel's isotopy stability theorem\cite{Handel:1985}, this means that every map in a pA mapping class is forced to have a topological entropy at least as large as that of the minimal map.  Most importantly, this is true of our flow map,
\begin{equation}
h\left( f^T \right) \geq h\left( f^{*}_{pA} \right).
\end{equation}
Thus, if we can force our mixing rods to move in a manner consistent with a pA braid, then the mixing as expressed by the topological entropy of the flow map, will have a well defined lower bound.

Through the Thurston-Nielsen classification theorem (TNCT)\cite{THURSTON:1988ty,Short:2000wi,BERNARDETE:1995ua}, every braid can be classified as being either finite-order (fo), pseudo-Anosov (pA), or reducible to a well defined combination of fo and pA components.  Finite-order braids have a minimal map with zero topological entropy, and therefore do not force any complexity on each map in the corresponding mapping class.  We include two fo braids in our analysis as a baseline for comparison to the pA braid.  pseudo-Anosov braids, as we have mentioned, force a lower bound on the topological entropy of every map in the corresponding mapping class.  We will refer to this lower bound as the pA topological entropy, $h\left( f^{*}_{pA} \right)$.  Except for a finite number of points, every point in a pA map, $f^{*}_{pA}$, has an expanding and contracting direction.  The common expansion factor, $\lambda_{pA}$, is related to the pA topological entropy by 
\begin{equation}
h\left( f^{*}_{pA} \right) = \ln \left(\lambda_{pA}\right).  
\end{equation}

The most general method for calculating the pA topological entropy uses the train-tracks algorithm of Bestvina and Handel \cite{BESTVINA:1995tx}, although an algorithm\cite{Moussafir} that makes use of Dynnikov coordinates\cite{Hall20091554} has shown promise\cite{Thiffeault:2010ir}.  We use a simpler method that takes advantage of the Burau representation\cite{bib:Kassel} of the braid group $B_3$.  Each braid generator can be represented by a $2\times2$ matrix,
\begin{equation}
\sigma_1 =  \begin{pmatrix}  1 & 1 \\ 0 & 1 \end{pmatrix} \; \; \; and \; \; \; 
\sigma_2 =  \begin{pmatrix}  1 & 0 \\ -1 & 1 \end{pmatrix}.
\end{equation}
These, along with their matrix inverses, can be used to create a matrix for every braid word.  For example, our pA braid, $\beta_{pA} = \left(\sigma_2^{-1} \sigma_1 \right)^3$, is represented by $\bigl( \begin{smallmatrix} 5&8\\ 8&13 \end{smallmatrix} \bigr)$.  The pA expansion factor, $\lambda_{pA}$, is given by the largest eigenvalue of this matrix.  For $\beta_{pA}$, this results in a value of $\lambda_{pA} = \phi^6$, where $\phi = \left(1+\sqrt{5}\right)/2$ is the golden ratio.  This ``golden braid" has a pA topological entropy per generator of 
\begin{equation}
h\left( f^{*}_{pA} \right)/N_g = \ln \left( \phi \right) = 0.4812,
\end{equation}
 which is the largest of any braid in $B_3$.

For each map in the mapping class of our pA braid, there exists\cite{Handel:1985} an invariant subset, $\mathbf{X}$, of the domain on which the dynamics are as complicated as that of the minimal representative.  It follows that fluid flow maps, $f^T$, induced by the pA stirring protocol have a topological entropy $h\left( f^T \right) \geq 6 \ln \left( \phi \right)$.  These maps will also be forced to have have local expansion factors which are larger than $\lambda_{pA}$ on some subset of $\mathbf{X}$ (though this connection is subtle).  Furthermore, this implies that the FTLE, which is directly related to the local expansion factors (see eq.~(\ref{NormTE})), is also bounded from below on some some subset of $\mathbf{X}$.  However, this theory says nothing about the size of $\mathbf{X}$.  Indeed, the area of $\mathbf{X}$ could be vanishingly small, in which case the dynamics on nearly the entire domain would not be forced to be complex.  This would physically mean that the expected mandatory mixing could have little effect on a real fluid.

\subsection{Numerical Experiment} \label{ssec:NumExp}

To numerically solve the Navier-Stokes equations in our domain, we used a finite element method (FEM), with the core of the code coming from the fenics project \cite{ans20553,bib:FEniCS}.  More specifically, we used a $100\times75$ rectangular crossed mesh ($3\times10^4$ elements) to spatially discretize the pressure field (linear scalar Lagrange elements) and velocity vector field (quadratic vector Lagrange elements).  The stability matrix, used for computation of the FTLE (see. eq.~(\ref{StabilityMatrix})) was also treated as a finite element tensor field (quadratic tensor Lagrange elements) and computed in tandem with the pressure and velocity fields.  A mesh fitted to the rods would be problematic, as the motion of the rods would exponentially stretch the mesh.  Our solution was to use a simple fixed cartesian mesh with a continuous immersed boundary method\cite{Lai2000705} to enforce the boundary conditions on the rods.  The boundary conditions were no-slip for both the outer walls and the rods.  We used Chorin's projection method\cite{Chorin} to discretize the equations of motion in time and decouple the pressure and velocity fields.  The time-step was chosen using a conservative CFL condition such that the rods take at least 10 time steps to cross the nearest neighbor element distance.  The Lagrangian quantities, the positions of particles passively advected by the velocity field and the Jacobian matrix of their orbits (see eq.~(\ref{JacobianEvolution})), were evolved using a fourth-order Runge-Kutta solver.  The orbits of these advected particles, initially at the center of each finite element, and the associated Jacobian matrices were then used to calculate the finite time Lyapunov exponent field (see eq.~(\ref{getFTLE1})).

The fenics FEM implementation of Corin's projection method for an incompressible Newtonian fluid has been tested on several benchmark flows\cite{bib:FEniCS} (see ch. 21), including lid-driven cavity flow, pressure-driven channel flow, flow past a cylinder, and Beltrami flow.  It was shown to be a good general solver with a nice balance of convergence properties and computational costs.  We did our own convergence study on grids ranging from $60\times45$ to $240\times180$ (all rectangular crossed grids) for both the final velocity field and the FTLE probability density function (pdf, see \ref{ssec::MixMeasure}), at both high and low Reynolds numbers.  The quantities generated with the finest grid were taken as a stand-in for the exact solution.  The velocity field error (FTLE pdf error) was calculated as the $\ell^2$-norm (absolute value) of the difference between approximate and exact values of the velocity vector field (FTLE pdf) at a few representative points in the domain (few FTLE values).  In all cases the order of convergence was between 1 and 2.  We chose the $100\times75$ rectangular crossed mesh since it was well within the range of asymptotic convergence and had a computational cost that fit within the time constraints of our study.  A relative error of roughly $0.02$ for the FTLE pdf calculated on the $100\times75$ rectangular crossed mesh (for most FTLE values) indicates that we can safely draw both quantitative and qualitative conclusions from our simulations.  While the main data represented in figures~(\ref{graph:FTLEMax}-\ref{graph:areaFTLE}) were generated on this grid, figures~(\ref{graph:FTLEpA1}-\ref{graph:pdfFTLE80}) were created using the $240\times180$ rectangular crossed grid to better highlight the fine structure.

For each simulation we chose a specific braid and Reynolds number.  We considered three braids (see fig.~(\ref{graph:Braidpic})), one pseudo-Anosov - $\beta_{pA}$ and two finite-order, full twist - $\beta_{\Delta^2}$ \& equivalent to the identity - $\beta_{Id}$.
\begin{subequations}
\begin{eqnarray}
\beta_{pA} &=& \sigma_2^{-1}\sigma_1\sigma_2^{-1}\sigma_1\sigma_2^{-1}\sigma_1 = \left( \sigma_2^{-1}\sigma_1 \right)^3 \\
\beta_{\Delta^2} &=& \sigma_2\sigma_1\sigma_2\sigma_1\sigma_2\sigma_1 =  \left( \sigma_2\sigma_1 \right)^3 \\
\beta_{Id} &=& \sigma_2\sigma_1\sigma_2\sigma_1^{-1}\sigma_2^{-1}\sigma_1
\end{eqnarray}
\end{subequations}
The pA braid was chosen because it has the largest topological entropy per generator, $\ln \phi$, of any three stranded braid.   This ``golden braid," should look familiar to those of you who have braided hair before.  The two finite-order braids are included to provide a baseline against which we compare the behavior of the pA braid.  Each braid has six generators, which alternate sides to ensure that the rods return to their original positions, and thus correspond to one period.  It is also important to note that for all three stirring protocols, the total length that the rods traverse is identical.  Therefore, at least for low Reynolds numbers, the work done by the rods on the fluid is the same for each braid.  Any mixing differences between the three braids will be due to topology alone.

For each braid we simulated a range of Reynolds numbers ($Re: \; 0.2-100$), with samples evenly spaced in $\log_{10}Re$ and double the sampling density near the center of this range.  We varied the kinematic viscosity to achieve this range, since $Re$ is calculated from the rod diameter, average rod speed, and kinematic viscosity.  While this range is squarely in the laminar regime, we will see that it is large enough to encompass an interesting range of behavior.

\subsection{Measuring the Mixing} \label{ssec::MixMeasure}

The existence of fluid mixing requires that material lines experience both stretching and folding\cite{Ottino1994}.  In our fluid, stretching is enforced by topology and folding is guaranteed by the boundedness of the fluid domain.  We would like to be able to quantify the amount of fluid mixing that our stirring rods create for each simulation.  While there are many different mixing measures to consider, we have taken the dynamic systems point of view\cite{ottino1989kinematics} in informing our choice of measure.  A global measure, such as the topological entropy of the corresponding particle advection map, does not help us deal with the question of what area participates in a high level of mixing.  For this we need a local measure, like the finite-time Lyapunov exponent (FTLE), $\Lambda \left( \mathbf{x}, T \right)$, which we can evaluate throughout the fluid domain. 

Roughly, the FTLE measures the maximum exponential rate of separation between an advected particle's trajectory and all other particle trajectories that started sufficiently close.  If the separation between two particles is $\left| \delta \mathbf{x}_0 \right|$ initially, then their separation after the simulation time $T$ is $\left| \delta \mathbf{x} \left(T \right) \right| \simeq \left|\delta \mathbf{x}_0\right| \exp\left( \Lambda T \right)$.  Thus, regions of our fluid that have a high FTLE will be greatly stretched out in the course of the stirring protocol.  This introduces fluid parcels to other, initially far away, fluid parcels, which enhances the mixing due to molecular diffusion.  

The finite-time Lyapunov exponent can be connected to other mixing ideas such as tracer gradients\cite{LapeyreFTLE2002},  advected particle concentration\cite{VothHallerGollubExperiment2002}, and the advection-diffusion equation \cite{FTLEReactionDiffusionTang1996, FTLEgeometryThiffeault2001}.  The FTLE has been used to characterize chaotic fluid flow via its power spectrum\cite{YuanOtt2000, ChaoticOrbitsPowerSpectrum} and quantify the performance of micro-mixers\cite{FTLEmicromixersSarkar2012}.  The FTLE has even been used to measure the degree of mixing in the Antarctic polar vortex through observational data\cite{GeophysicsFTLEBowman1993} and predict the rate of chemical reactions in experiments\cite{GollubPhysRevLett2006}.

To make the intuitive notion of a FTLE more concrete and computable, consider the Jacobian matrix  $\mathbf{J}^t_{ij}\left(\mathbf{x}_0\right) = \partial x_i\left(t\right)/\partial x_j \left(0\right)$, which transports linearized neighborhoods under the fluid flow, $\delta \mathbf{x} \left(t \right) = \mathbf{J}^t\left( \mathbf{x}_0 \right) \delta \mathbf{x}_0$.  One can then use the maximum eigenvalue, $\lambda_{max}$, of $\mathbf{J}^T$ at each point, $\mathbf{x}_0$,  to calculate the FTLE field,
\begin{equation}
\Lambda\left( \mathbf{x}_0, T \right) = \frac{1}{T} \ln\left( \lambda_{max}\left( \mathbf{J}^T\left( \mathbf{x}_0 \right) \right) \right).\label{FTLEcalc1}
\end{equation}
To find $\mathbf{J}^T\left( \mathbf{x}_0 \right)$, integrate 
\begin{equation}
d \mathbf{J}^t\left( \mathbf{x}_0 \right)/dt = \mathbf{A}\left(\mathbf{x}\left( t \right)\right) \mathbf{J}^t\left( \mathbf{x}_0 \right) \label{JacobianEvolution}
\end{equation}
 along the trajectory, $\mathbf{x}\left( t \right)$, of each advected particle, starting with $\mathbf{J}^0 \left( \mathbf{x}_0 \right) = \mathbf{1}$.  Here, 
 \begin{equation}
A_{ij}\left(\mathbf{x}\right) = \partial \dot{x}_i\left( \mathbf{x} \right)/\partial x_j \label{StabilityMatrix}
 \end{equation}
 is the stability matrix, which describes the linear shearing on a neighborhood of $\mathbf{x}$ due the fluid flow\cite{ChaosBook}.  

Since the Jacobian has the semi-group property, we can split up the Jacobian and express the FTLE as
\begin{equation}
\Lambda\left( \mathbf{x}_0, T, \hat{u}_0 \right) = \frac{1}{T} \ln \left\Vert \left( \prod_{i=1}^{N} \mathbf{J}^{\Delta t}\left( \mathbf{x}\left( T - i \Delta t \right) \right) \right) \hat{u}_0 \right\Vert, \label{FTLEcalc2}
\end{equation}
where $N \Delta t = T$ and $\hat{u}_0$ is a unit vector.  Then $\Lambda\left( \mathbf{x}_0,T \right)$ is found by maximizing over all directions of $\hat{u}_0$.  In practice, most all $\hat{u}_0$ will quickly converge, through the action of $\mathbf{J}^{\Delta t}$, to the maximum stretching direction.  Therefore, $\Lambda\left( \mathbf{x}_0, T,\hat{u}_0 \right)$ is fairly insensitive to $\hat{u}_0$, and we can pick a convenient vector.  We can simplify further to arrive at
\begin{equation}
\Lambda\left( \mathbf{x}_0, T \right) = \frac{1}{T} \sum_{i = 0}^{N-1} \ln \left\Vert \vec{S}_i \right\Vert, \label{getFTLE1}
\end{equation}
where 
\begin{equation}
\vec{S}_i = \mathbf{J}^{\Delta t}\left( \mathbf{x}\left( i \Delta t \right) \right)\hat{u}_i, \;\;\;  \hat{u}_{i} = \vec{S}_{i-1}/ \left\| \vec{S}_{i-1} \right\|, \label{getFTLE2}
\end{equation}
and $\hat{u}_0$ is chosen in some consistent, but unimportant, direction\cite{Benettin76}.  This expression suggests calculating $\mathbf{J}^{\Delta t}\left( \mathbf{x}\left( i \Delta t \right) \right)$ and $\vec{S}_i$ concurrently with each particle trajectory, and keeping a running sum of $\ln \| \vec{S}_i \|$.  The periodic normalization, orthonormalization if keeping track of the minimum FTLE as well\cite{Shimada79}, obviates the need to find the spectrum of an overall Jacobian that might be ill-conditioned.

\begin{figure}[htbp]
 \centering
  \large
   \includegraphics{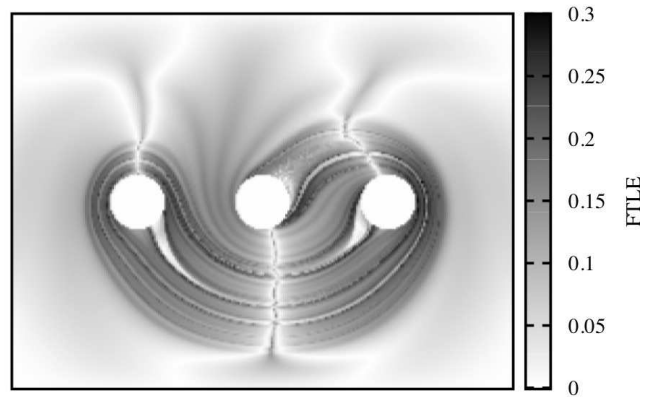}
  \normalsize
    \caption{FTLE field for the pA stirring protocol at a Reynolds number of $Re = 0.2$.}
    \label{graph:FTLEpA1}
\end{figure}

\begin{figure*}[htbp]
 \centering
  \large
  \includegraphics{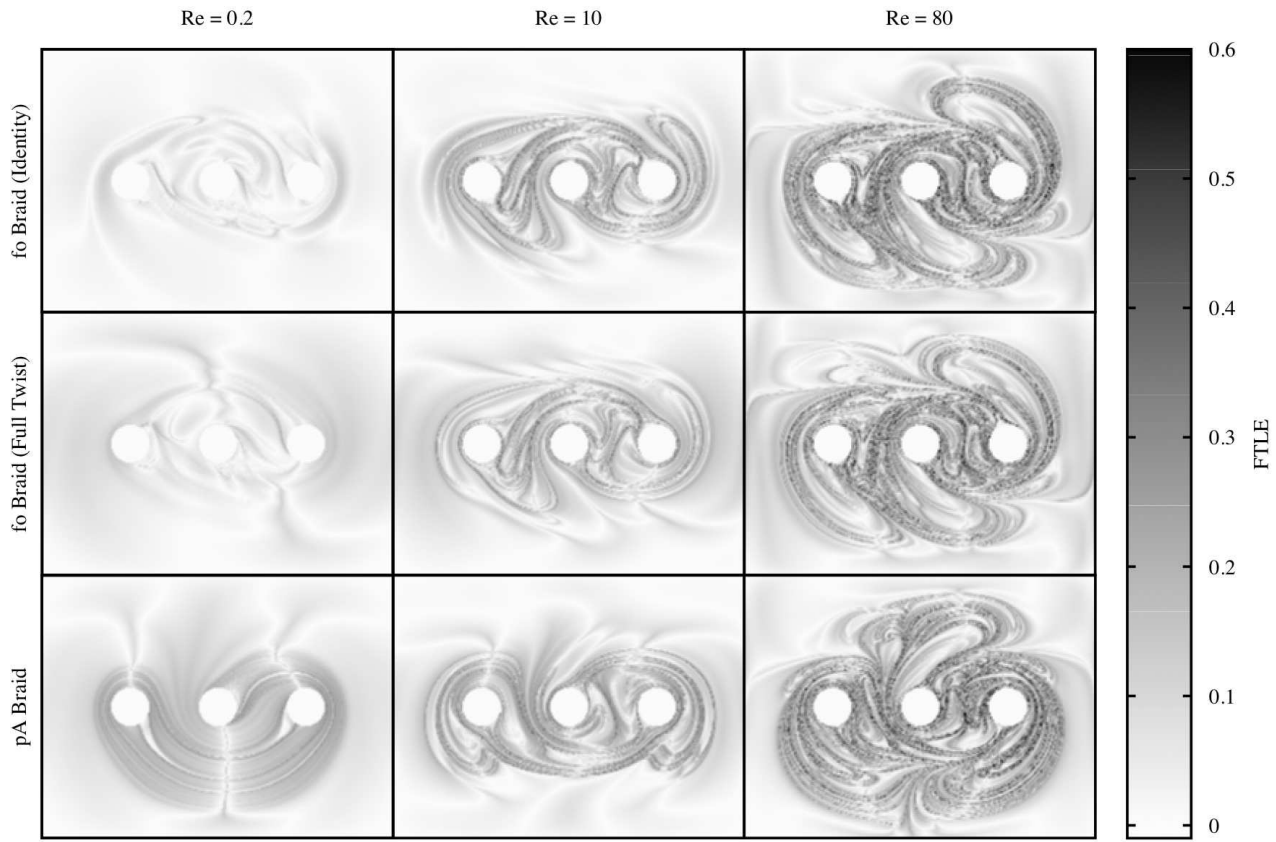}
  \normalsize
    \caption{FTLE fields.  Rows (top to bottom): finite-order braid (isotopic to identity), finite-order braid (full twist), and pseudo-Anosov ``golden" braid.  Columns (left to right): Reynolds number $Re = 0.2$, $Re = 10$, and $Re = 80$.}
    \label{graph:FTLEall}
\end{figure*}

With this algorithm, we computed the FTLE at the center of each lattice element, resulting in a good approximation of the FTLE map, $\Lambda\left( \mathbf{x}_0,T \right)$ (e.g. see fig.~(\ref{graph:FTLEpA1}) or fig.~(\ref{graph:FTLEall})).  In order to capture the entirety of the motion and make the FTLE comparable to the pA topological entropy (e.g. see Eq.~(\ref{NormTE})), the time, T, used in the computation of the FTLE was chosen to be equal to the stirring protocol time of $6\pi$ seconds.  Note that these maps give fairly intuitive information about where the mixing is occurring.

To compare the effect of different braids and different Reynolds numbers on mixing, we needed to extract from each FTLE field a few simple measures.  As a partial step in this direction, consider the FTLE probability density function (PDF), $\rho\left( \Lambda \right)$, for each FTLE map, where the probability measure is the fractional area.  Thus the fraction of the total fluid domain that has FTLE between $\Lambda_1$ and $\Lambda_2$ is given by $\int^{\Lambda2}_{\Lambda1} \rho\left( \Lambda \right) d \Lambda$.  The PDF of each FTLE map provides useful information in its own right, e.g. see figures~(\ref{graph:pdfFTLE}-\ref{graph:pdfFTLE80}), and allows us a convenient way to define three useful measures of mixing.

The first is the maximum FTLE, $\Lambda_{max}$, which is calculated as the FTLE value for which $99\%$ of the fluid by area has a smaller FTLE value,
\begin{equation}
 \int^{\Lambda_{max}}_0  \rho\left( \Lambda \right) d \Lambda = 0.99. \label{Lmax}
\end{equation}
The value of $0.99$ is used instead of something closer to $1$, to ensure that the finite grid size does not result in widely fluctuating values of $\Lambda_{max}$ across different Reynolds numbers.  Thus, the actual maximum value of the FTLE map, given by the first FTLE value, $\Lambda^{*}_{max}$, such that $ \int^{\Lambda^{*}_{max}}_0  \rho\left( \Lambda \right) d \Lambda = 1$, will be slightly larger than $\Lambda_{max}$.  As we will see, the minimal mixing mandated by the TN theory will show up most clearly in a plot of the maximum FTLE.  

The second measure is the area-averaged FTLE, $\widetilde{\Lambda}$, and is simply the first moment of the PDF,
\begin{equation}
\widetilde{\Lambda} =  \int^{\Lambda^{*}_{max}}_0  \Lambda \rho\left( \Lambda \right) d \Lambda. \label{Lavg}
\end{equation}
It provides a way to track the level of mixing in the majority of the fluid, and is therefore useful for addressing how much of the fluid is compelled to have a high level of mixing by the pA stirring protocol .

The final measure is the fractional area in which the FTLE value is larger than the normalized pA topological entropy, $A_{pA}$.  To compare the FTLE values with the pA topological entropy, we must first normalize the latter.  Both the pA topological entropy  and the FTLE are given as the logarithm of a stretching factor, $\ln \left( \lambda_{pA} \right)$ and $\ln \left( \lambda_{max} \right)/T$ respectively.  Since we should be able to compare the two stretching factors, we need to divide the pA topological entropy by the total time, $T = 6\pi$, in order to make meaningful comparisons to FTLE values.  Thus, the normalized pA topological entropy is
\begin{equation}
\Lambda_{pA} \equiv h\left( f^{*}_{pA} \right)/T = \frac{1}{\pi}\ln \phi = 0.1532. \label{NormTE}
\end{equation}
Now the final measure, $A_{pA}$, is defined as
\begin{equation}
A_{pA} =  \int^{\Lambda^{*}_{max}}_{\Lambda_{pA}}  \rho\left( \Lambda \right) d \Lambda. \label{areapA}
\end{equation}

With the max FTLE, area-averaged FTLE, and fractional area measure in hand, we can now compare them across different Reynolds numbers and different braids.

\section{Results}

\begin{figure}[htbp]
 \centering
  \large
   \includegraphics{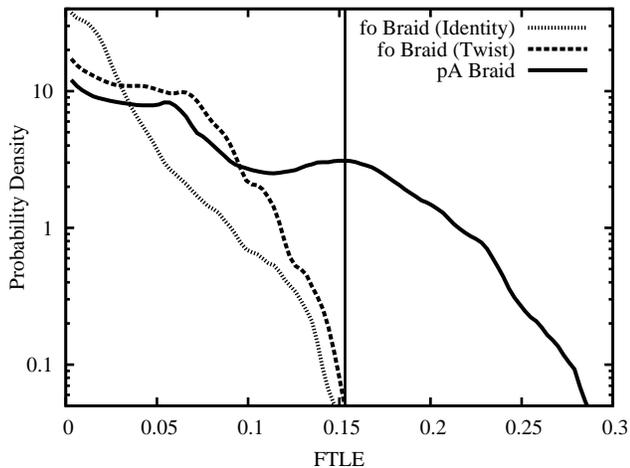}
  \normalsize
    \caption{The FTLE probability density function (PDF) at $Re = 0.2$, plotted in log-scale, for each braiding protocol.  The normalized pA topological entropy, $\Lambda_{pA} = 0.1532$ -  Eq.~(\ref{NormTE}), is shown as a vertical line for reference.}
    \label{graph:pdfFTLE}
\end{figure}

Consider the collection of FTLE maps in fig.~(\ref{graph:FTLEall}).  Just by visual inspection, it is apparent that at the lowest Reynolds number, $Re = 0.2$, the pA stirring protocol has an appreciably higher level of mixing over a larger area than either of the finite-order stirring protocols.  The main reason for this difference is topology, as we can see in the FTLE probability density, fig.~(\ref{graph:pdfFTLE}), for each braiding protocol at this Reynolds number.  Notice that at a FTLE value equal to the normalized pA topological entropy ($\Lambda_{pA} = 0.1532$), there is a ``bump" in the pA PDF caused by the topologically mandated exponential stretching of material lines.  

The other bump at $\Lambda \simeq 0.06$ is likely due to sub-exponential stretching associated with fluid outside the central mixing region.  When the simulation was run for twice as long (pA braid: $\left( \sigma_2^{-1}\sigma_1 \right)^6$) this bump was still evident in the PDF, though at a reduced FTLE value of $\Lambda \simeq 0.04$.  If this stretching is linear, then repeated application of the fluid advection map would cause the FTLE value of this bump to asymptotically approach zero as $\Lambda \simeq \ln\left(n\right)/nT$ for $n$ compositions of the advection map.  The absence of this $\Lambda \simeq 0.06$ bump for the identity finite-order stirring protocol reflects the fact that on each side of the stirring rod ensemble there are both CW and CCW rod movements, unlike the other two stirring protocols.  Thus, in the Stokes regime the rods will stir the fluid outside the central mixing region in competing directions over the course of the identity finite-order protocol, which reduces the rate of linear stretching.

\begin{figure}[htbp]
 \centering
  \large
   \includegraphics{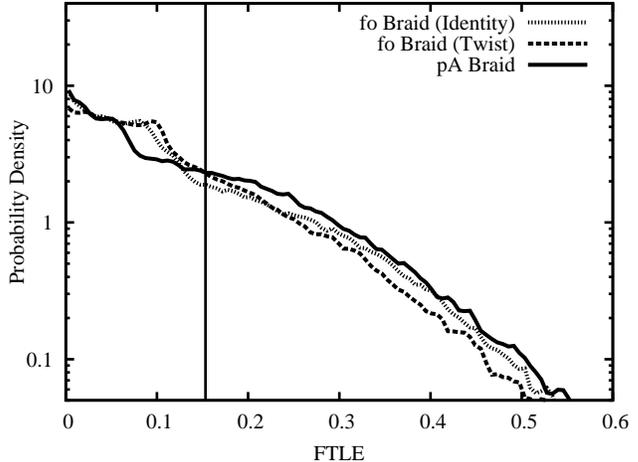}
  \normalsize
    \caption{The FTLE probability density function (PDF) at $Re = 80$, plotted in log-scale, for each braiding protocol.  The normalized pA topological entropy, $\Lambda_{pA} = 0.1532$ -  Eq.~(\ref{NormTE}), is shown as a vertical line for reference.  Note the difference in FTLE range as compared to fig.~(\ref{graph:pdfFTLE}).}
    \label{graph:pdfFTLE80}
\end{figure}

The size of the $\Lambda_{pA} = 0.1532$, exponential stretching, bump is largest in the Stokes regime, where a large quantity of fluid is entrained with the stirring rods.  Indeed, at higher Reynolds numbers, the PDFs for all three stirring protocols begin to converge (see fig.~(\ref{graph:pdfFTLE80})).  This can likewise be seen in fig.~(\ref{graph:FTLEall}), where the difference between the three protocols is less visible for the two higher Reynolds numbers.  To uncover the range of Reynolds numbers over which the pA stirring protocol retains its advantage, we must look at the individual measures for each FTLE map, starting with the maximum FTLE.

\begin{figure}[htbp]
 \centering
  \large
   \includegraphics{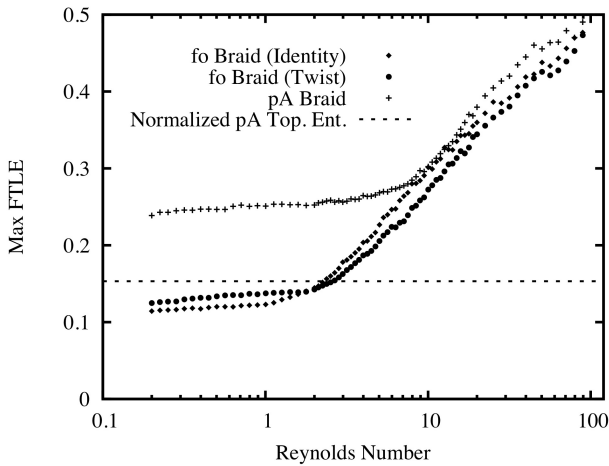}
  \normalsize
    \caption{The maximum FTLE,  $\Lambda_{max}$ - Eq.~(\ref{Lmax}), of all three braids for a range of Reynolds numbers.  The normalized pA topological entropy, $\Lambda_{pA} = 0.1532$ - Eq.~(\ref{NormTE}), is shown for comparison.}
    \label{graph:FTLEMax}
\end{figure}

The maximum FTLE,  $\Lambda_{max}$, over the Reynolds number range is plotted in fig.~(\ref{graph:FTLEMax}).  It clearly shows that while the pA stirring protocol has a larger $\Lambda_{max}$ value than that of either finite-order protocol over the entire $Re$ range, there are distinct regions based on the size of this gap.  

For \mbox{$Re$ {\small$\lesssim$} $2$}, where this $\Lambda_{max}$ gap between pA and fo protocols is largest, we see only a very weak Reynolds number dependence.  This insensitivity to Reynolds number also shows up in a comparison of FTLE PDFs, and is reflective of the fact that the fluid is in the Stokes flow regime.  In this regime, topological differences have their largest effect on mixing.  Indeed, the normalized pA topological entropy, shown in fig.~(\ref{graph:FTLEMax}), is a lower bound on $\Lambda_{max}$ for the pA protocol but not for either of the two finite-order protocols.  That it is not a tight lower bound simply reflects that along a closed material line which is forced to have the proscribed topological stretching in aggregate, there will be sections with both larger and smaller local stretching.  This roughly accounts for the spread in FTLE about the ``bump" in probability density shown in fig.~(\ref{graph:pdfFTLE}).

For slightly larger Reynolds number, there is a transitional region, \mbox{$2$ {\small$\lesssim$} $Re$ {\small$\lesssim$} $8$}, where the increasingly important inertia of the fluid causes the $\Lambda_{max}$ of the finite-order protocols to rise.  The pA protocol has a $\Lambda_{max}$ that remains flat in this region, indicating that the lower bound on mixing set by the topology is still large enough to dictate the maximum mixing.

For the region $Re$ {\small$\gtrsim$} $8$, all three protocols behave similarly, and the $\Lambda_{max}$ is dictated not by topology, but by the inertial motion of the fluid.  As a brief aside, note that in this regime, $\Lambda_{max}$ appears to scale linearly with $\log Re$.  Many studies\cite{micromixerReynoldsXia2006, ChaoticAdvecInertiaWang2009, KinicsMixerReDepHobbs1998, PanetaryMixerClifford2004}, both experimental and numerical, have shown that mixing due to chaotic advection increases with $Re$ in the low Reynolds number, near-Stokes, regime.  Interestingly, mixing due to chaotic advection can actually decreases at high Re in the turbulent regime\cite{2001Mezic, LEvReTurbBaggaley2009}.


\begin{figure}[htbp]
 \centering
  \large
   \includegraphics{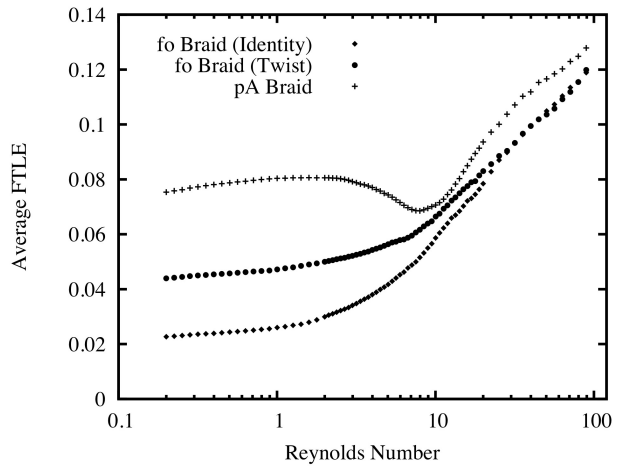}
  \normalsize
    \caption{The area averaged FTLE, $\widetilde{\Lambda}$ - Eq.~(\ref{Lavg}), of all three braids for a range of Reynolds numbers.}
    \label{graph:FTLEavg}
\end{figure}

So far we have seen that the topological lower bound on mixing is useful for describing the level at which maximum mixing occurs when the fluid is in the Stokes regime.  However, to get a sense of the size of the fluid area that participates in a high level of mixing, we must look at the behavior of the two remaining measures.  First, consider the area averaged FTLE, $\widetilde{\Lambda}$, in fig.~(\ref{graph:FTLEavg}).  While it does not directly measure area, $\widetilde{\Lambda}$ is larger for fluids that have a larger fraction of their area participating in a high level of mixing.

In the \mbox{$Re$ {\small$\lesssim$} $2$} regime, we see that the topology of the pA stirring protocol forces $\widetilde{\Lambda}$ to be larger than it is in the finite-order protocols.  Thus, we can put to rest any worry that the mandated high level of mixing only occurs on a vanishingly small area.  In the transitional region, \mbox{$2$ {\small$\lesssim$} $Re$ {\small$\lesssim$} $8$}, the area averaged FTLE actually decreases for the pA protocol.  At first this might seem unexpected, as the same is not true of the two finite-order protocols, and $\Lambda_{max}$ did not decrease for the pA protocol.  However, as the Reynolds number increases through this region, less and less of the fluid is viscously entrained with the stirring rods, and therefore less of the area is forced to have the minimal mixing.  Eventually, for high enough Reynolds number, \mbox{$Re$ {\small$\gtrsim$} $8$}, dynamic mixing due to the greater importance of fluid inertia is the dominant source of mixing.

\begin{figure}[htbp]
 \centering
  \large
   \includegraphics{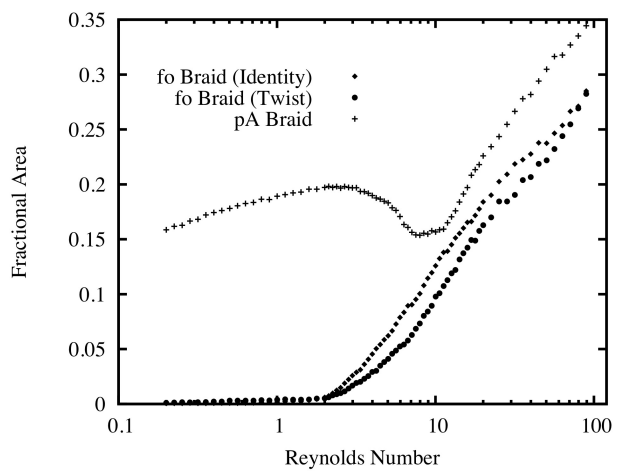}
  \normalsize
    \caption{The fractional area, $A_{pA}$ - Eq.~(\ref{areapA}), in which the FTLE is larger than the normalized pA topological entropy, $\Lambda_{pA} = 0.1532$ - Eq.~(\ref{NormTE}).  This fractional area is plotted for all three braids and a range of Reynolds numbers.}
    \label{graph:areaFTLE}
\end{figure}

Finally, we consider directly the area affected by the pA stirring protocol through a plot, fig.~(\ref{graph:areaFTLE}), of the fractional area,  $A_{pA}$, over which the FTLE is greater than the normalized topological entropy, $\Lambda_{pA}$.  This shows that in the Stokes regime, between $1/6$ and $1/5$ of the fluid mixes at or above the level enforced by topology.  This is larger than the area of the fluid which is in the direct path of the moving rods and smaller than the area inclosed by this path.  We also see a dip in the fractional area in the \mbox{$2$ {\small$\lesssim$} $Re$ {\small$\lesssim$} $8$} range, which follows from the notion that with increasing Reynolds number, less of the fluid is entrained with the stirring rods.  While the area of topologically enforced mixing decreases for \mbox{$Re$ {\small$\gtrsim$} $2$}, it is important to note that the pA stirring protocol still has higher values of $\Lambda_{max}$, $\widetilde{\Lambda}$, and $A_{pA}$, albeit at a much reduced advantage with respect to the finite-order protocols.

\section{Conclusion}

Topological chaos guarantees that the advection map associated with a pseudo-Anosov stirring protocol will have a topological entropy at least as large as that of the minimal representative in the corresponding mapping class.  This topologically imposed lower bound on the mixing could potentially be both very loose, and supported on a vanishingly small fraction of the fluid domain.  Fortunately, neither of these issues are problematic in the Stokes regime.  For \mbox{$Re$ {\small$\lesssim$} $2$}, the gap between the maximum FTLE of the pA and finite-order protocols is larger than for any other Reynolds number range tested, and consistent with the topological lower bound.  The two area related measures, the area averaged FTLE and the fractional area with a FTLE greater than the normalized pA topological entropy, indicate that the area of topologically enforced mixing is comparable with the area enclosed by the stirring rods.  

However, at higher Reynolds numbers, \mbox{$Re$ {\small$\gtrsim$} $8$}, topologically enforced mixing is overshadowed by the dynamic mixing due to the increased importance of fluid inertia.  In this range, the differences between the three stirring protocols are minimal.  Topological mixing still enforces larger max and average FTLE values for the pA protocol compared to the finite-order protocols, however the difference becomes an increasingly smaller fraction of the total.

In the transitional region, \mbox{$2$ {\small$\lesssim$} $Re$ {\small$\lesssim$} $8$}, a somewhat unexpected phenomenon occurs.  Before the dynamic mixing becomes dominant, the area of support for topological mixing decreases.  This is reflected in the dip in average FTLE and fractional area, $A_{pA}$, and results when less fluid is entrained with the string rods due to decreased viscosity.

Overall, these results affirm that topological chaos is best suited for investigating fluid mixing in the Stokes regime.  For higher Reynolds numbers, the area of topologically enforced mixing decreases, and is eventually overcome by dynamic mixing associated with fluid inertia.

\bibliography{TopMix}

\begin{thebibliography}{52}%
\makeatletter
\providecommand \@ifxundefined [1]{%
 \@ifx{#1\undefined}
}%
\providecommand \@ifnum [1]{%
 \ifnum #1\expandafter \@firstoftwo
 \else \expandafter \@secondoftwo
 \fi
}%
\providecommand \@ifx [1]{%
 \ifx #1\expandafter \@firstoftwo
 \else \expandafter \@secondoftwo
 \fi
}%
\providecommand \natexlab [1]{#1}%
\providecommand \enquote  [1]{``#1''}%
\providecommand \bibnamefont  [1]{#1}%
\providecommand \bibfnamefont [1]{#1}%
\providecommand \citenamefont [1]{#1}%
\providecommand \href@noop [0]{\@secondoftwo}%
\providecommand \href [0]{\begingroup \@sanitize@url \@href}%
\providecommand \@href[1]{\@@startlink{#1}\@@href}%
\providecommand \@@href[1]{\endgroup#1\@@endlink}%
\providecommand \@sanitize@url [0]{\catcode `\\12\catcode `\$12\catcode
  `\&12\catcode `\#12\catcode `\^12\catcode `\_12\catcode `\%12\relax}%
\providecommand \@@startlink[1]{}%
\providecommand \@@endlink[0]{}%
\providecommand \url  [0]{\begingroup\@sanitize@url \@url }%
\providecommand \@url [1]{\endgroup\@href {#1}{\urlprefix }}%
\providecommand \urlprefix  [0]{URL }%
\providecommand \Eprint [0]{\href }%
\providecommand \doibase [0]{http://dx.doi.org/}%
\providecommand \selectlanguage [0]{\@gobble}%
\providecommand \bibinfo  [0]{\@secondoftwo}%
\providecommand \bibfield  [0]{\@secondoftwo}%
\providecommand \translation [1]{[#1]}%
\providecommand \BibitemOpen [0]{}%
\providecommand \bibitemStop [0]{}%
\providecommand \bibitemNoStop [0]{.\EOS\space}%
\providecommand \EOS [0]{\spacefactor3000\relax}%
\providecommand \BibitemShut  [1]{\csname bibitem#1\endcsname}%
\let\auto@bib@innerbib\@empty
\bibitem [{\citenamefont {Finn}, \citenamefont {Cox},\ and\ \citenamefont
  {Byrne}(2004)}]{FinnCoxMix}%
  \BibitemOpen
  \bibfield  {author} {\bibinfo {author} {\bibfnamefont {M.}~\bibnamefont
  {Finn}}, \bibinfo {author} {\bibfnamefont {S.}~\bibnamefont {Cox}}, \ and\
  \bibinfo {author} {\bibfnamefont {H.}~\bibnamefont {Byrne}},\ }\bibfield
  {title} {\enquote {\bibinfo {title} {Mixing measures for a two-dimensional
  chaotic stokes flow},}\ }\href@noop {} {\bibfield  {journal} {\bibinfo
  {journal} {Journal of Engineering Mathematics}\ }\textbf {\bibinfo {volume}
  {48}},\ \bibinfo {pages} {129--155} (\bibinfo {year} {2004})}\BibitemShut
  {NoStop}%
\bibitem [{\citenamefont {{Aref}}\ \emph {et~al.}(2014)\citenamefont {{Aref}},
  \citenamefont {{Blake}}, \citenamefont {{Budi{\v s}i{\'c}}}, \citenamefont
  {{Cartwright}}, \citenamefont {{Clercx}}, \citenamefont {{Feudel}},
  \citenamefont {{Golestanian}}, \citenamefont {{Gouillart}}, \citenamefont
  {{Le Guer}}, \citenamefont {{van Heijst}}, \citenamefont {{Krasnopolskaya}},
  \citenamefont {{MacKay}}, \citenamefont {{Meleshko}}, \citenamefont
  {{Metcalfe}}, \citenamefont {{Mezi{\'c}}}, \citenamefont {{de Moura}},
  \citenamefont {{El Omari}}, \citenamefont {{Piro}}, \citenamefont
  {{Speetjens}}, \citenamefont {{Sturman}}, \citenamefont {{Thiffeault}},\ and\
  \citenamefont {{Tuval}}}]{2014arXiv1403.2953A}%
  \BibitemOpen
  \bibfield  {author} {\bibinfo {author} {\bibfnamefont {H.}~\bibnamefont
  {{Aref}}}, \bibinfo {author} {\bibfnamefont {J.~R.}\ \bibnamefont {{Blake}}},
  \bibinfo {author} {\bibfnamefont {M.}~\bibnamefont {{Budi{\v s}i{\'c}}}},
  \bibinfo {author} {\bibfnamefont {J.~H.~E.}\ \bibnamefont {{Cartwright}}},
  \bibinfo {author} {\bibfnamefont {H.~J.~H.}\ \bibnamefont {{Clercx}}},
  \bibinfo {author} {\bibfnamefont {U.}~\bibnamefont {{Feudel}}}, \bibinfo
  {author} {\bibfnamefont {R.}~\bibnamefont {{Golestanian}}}, \bibinfo {author}
  {\bibfnamefont {E.}~\bibnamefont {{Gouillart}}}, \bibinfo {author}
  {\bibfnamefont {Y.}~\bibnamefont {{Le Guer}}}, \bibinfo {author}
  {\bibfnamefont {G.~F.}\ \bibnamefont {{van Heijst}}}, \bibinfo {author}
  {\bibfnamefont {T.~S.}\ \bibnamefont {{Krasnopolskaya}}}, \bibinfo {author}
  {\bibfnamefont {R.~S.}\ \bibnamefont {{MacKay}}}, \bibinfo {author}
  {\bibfnamefont {V.~V.}\ \bibnamefont {{Meleshko}}}, \bibinfo {author}
  {\bibfnamefont {G.}~\bibnamefont {{Metcalfe}}}, \bibinfo {author}
  {\bibfnamefont {I.}~\bibnamefont {{Mezi{\'c}}}}, \bibinfo {author}
  {\bibfnamefont {A.~P.~S.}\ \bibnamefont {{de Moura}}}, \bibinfo {author}
  {\bibfnamefont {K.}~\bibnamefont {{El Omari}}}, \bibinfo {author}
  {\bibfnamefont {O.}~\bibnamefont {{Piro}}}, \bibinfo {author} {\bibfnamefont
  {M.~F.~M.}\ \bibnamefont {{Speetjens}}}, \bibinfo {author} {\bibfnamefont
  {R.}~\bibnamefont {{Sturman}}}, \bibinfo {author} {\bibfnamefont {J.-L.}\
  \bibnamefont {{Thiffeault}}}, \ and\ \bibinfo {author} {\bibfnamefont
  {I.}~\bibnamefont {{Tuval}}},\ }\bibfield  {title} {\enquote {\bibinfo
  {title} {{Frontiers of chaotic advection}},}\ }\href@noop {} {\bibfield
  {journal} {\bibinfo  {journal} {ArXiv e-prints}\ } (\bibinfo {year}
  {2014})},\ \Eprint {http://arxiv.org/abs/1403.2953} {arXiv:1403.2953
  [nlin.CD]} \BibitemShut {NoStop}%
\bibitem [{\citenamefont {Samelson}(2013)}]{Samelson13}%
  \BibitemOpen
  \bibfield  {author} {\bibinfo {author} {\bibfnamefont {R.}~\bibnamefont
  {Samelson}},\ }\bibfield  {title} {\enquote {\bibinfo {title} {Lagrangian
  motion, coherent structures, and lines of persistent material strain},}\
  }\href@noop {} {\bibfield  {journal} {\bibinfo  {journal} {Annual Review of
  Marine Science}\ }\textbf {\bibinfo {volume} {5}},\ \bibinfo {pages}
  {137--163} (\bibinfo {year} {2013})}\BibitemShut {NoStop}%
\bibitem [{\citenamefont {Thiffeault}(2010)}]{Thiffeault:2010ir}%
  \BibitemOpen
  \bibfield  {author} {\bibinfo {author} {\bibfnamefont {J.-L.}\ \bibnamefont
  {Thiffeault}},\ }\bibfield  {title} {\enquote {\bibinfo {title} {{Braids of
  entangled particle trajectories}},}\ }\href@noop {} {\bibfield  {journal}
  {\bibinfo  {journal} {AIP Chaos}\ }\textbf {\bibinfo {volume} {20}},\
  \bibinfo {pages} {017516} (\bibinfo {year} {2010})}\BibitemShut {NoStop}%
\bibitem [{\citenamefont {Aref}(1984)}]{Aref:1984}%
  \BibitemOpen
  \bibfield  {author} {\bibinfo {author} {\bibfnamefont {H.}~\bibnamefont
  {Aref}},\ }\bibfield  {title} {\enquote {\bibinfo {title} {Stirring by
  chaotic advection},}\ }\href {\doibase 10.1017/S0022112084001233} {\bibfield
  {journal} {\bibinfo  {journal} {Journal of Fluid Mechanics}\ }\textbf
  {\bibinfo {volume} {143}},\ \bibinfo {pages} {1--21} (\bibinfo {year}
  {1984})}\BibitemShut {NoStop}%
\bibitem [{\citenamefont {Aref}(2002)}]{bib:Aref1}%
  \BibitemOpen
  \bibfield  {author} {\bibinfo {author} {\bibfnamefont {H.}~\bibnamefont
  {Aref}},\ }\bibfield  {title} {\enquote {\bibinfo {title} {The development of
  chaotic advection},}\ }\href@noop {} {\bibfield  {journal} {\bibinfo
  {journal} {Phys. Fluids}\ }\textbf {\bibinfo {volume} {14}},\ \bibinfo
  {pages} {1315--1325} (\bibinfo {year} {2002})}\BibitemShut {NoStop}%
\bibitem [{\citenamefont {Boyland}, \citenamefont {Aref},\ and\ \citenamefont
  {Stremler}(2000)}]{Boyland:2000uc}%
  \BibitemOpen
  \bibfield  {author} {\bibinfo {author} {\bibfnamefont {P.}~\bibnamefont
  {Boyland}}, \bibinfo {author} {\bibfnamefont {H.}~\bibnamefont {Aref}}, \
  and\ \bibinfo {author} {\bibfnamefont {M.}~\bibnamefont {Stremler}},\
  }\bibfield  {title} {\enquote {\bibinfo {title} {Topological fluid mechanics
  of stirring},}\ }\href@noop {} {\bibfield  {journal} {\bibinfo  {journal} {J.
  Fluid Mech.}\ }\textbf {\bibinfo {volume} {403}},\ \bibinfo {pages}
  {277--304} (\bibinfo {year} {2000})}\BibitemShut {NoStop}%
\bibitem [{\citenamefont {Boyland}(1994)}]{Boyland:1994ud}%
  \BibitemOpen
  \bibfield  {author} {\bibinfo {author} {\bibfnamefont {P.}~\bibnamefont
  {Boyland}},\ }\bibfield  {title} {\enquote {\bibinfo {title} {{Topological
  Methods in Surface Dynamics}},}\ }\href@noop {} {\bibfield  {journal}
  {\bibinfo  {journal} {Topology And Its Applications}\ }\textbf {\bibinfo
  {volume} {58}},\ \bibinfo {pages} {223--298} (\bibinfo {year}
  {1994})}\BibitemShut {NoStop}%
\bibitem [{\citenamefont {Thiffeault}\ and\ \citenamefont
  {Finn}(2006)}]{Thiffeault:2006jp}%
  \BibitemOpen
  \bibfield  {author} {\bibinfo {author} {\bibfnamefont {J.-L.}\ \bibnamefont
  {Thiffeault}}\ and\ \bibinfo {author} {\bibfnamefont {M.~D.}\ \bibnamefont
  {Finn}},\ }\bibfield  {title} {\enquote {\bibinfo {title} {{Topology, braids
  and mixing in fluids}},}\ }\href@noop {} {\bibfield  {journal} {\bibinfo
  {journal} {Philosophical Transactions Of The Royal Society A-Mathematical
  Physical And Engineering Sciences}\ }\textbf {\bibinfo {volume} {364}},\
  \bibinfo {pages} {3251--3266} (\bibinfo {year} {2006})}\BibitemShut {NoStop}%
\bibitem [{\citenamefont {Thiffeault}(2005)}]{Thiffeault:2005hn}%
  \BibitemOpen
  \bibfield  {author} {\bibinfo {author} {\bibfnamefont {J.}~\bibnamefont
  {Thiffeault}},\ }\bibfield  {title} {\enquote {\bibinfo {title} {{Measuring
  topological chaos}},}\ }\href@noop {} {\bibfield  {journal} {\bibinfo
  {journal} {Phys. Rev. Lett.}\ }\textbf {\bibinfo {volume} {94}},\ \bibinfo
  {pages} {084502} (\bibinfo {year} {2005})}\BibitemShut {NoStop}%
\bibitem [{\citenamefont {Thiffeault}\ \emph {et~al.}(2008)\citenamefont
  {Thiffeault}, \citenamefont {Finn}, \citenamefont {Gouillart},\ and\
  \citenamefont {Hall}}]{Thiffeault:2008er}%
  \BibitemOpen
  \bibfield  {author} {\bibinfo {author} {\bibfnamefont {J.-L.}\ \bibnamefont
  {Thiffeault}}, \bibinfo {author} {\bibfnamefont {M.~D.}\ \bibnamefont
  {Finn}}, \bibinfo {author} {\bibfnamefont {E.}~\bibnamefont {Gouillart}}, \
  and\ \bibinfo {author} {\bibfnamefont {T.}~\bibnamefont {Hall}},\ }\bibfield
  {title} {\enquote {\bibinfo {title} {{Topology of chaotic mixing
  patterns}},}\ }\href@noop {} {\bibfield  {journal} {\bibinfo  {journal} {AIP
  Chaos}\ }\textbf {\bibinfo {volume} {18}},\ \bibinfo {pages} {033123}
  (\bibinfo {year} {2008})}\BibitemShut {NoStop}%
\bibitem [{\citenamefont {Finn}, \citenamefont {Thiffeault},\ and\
  \citenamefont {Gouillart}(2006)}]{Finn200692}%
  \BibitemOpen
  \bibfield  {author} {\bibinfo {author} {\bibfnamefont {M.~D.}\ \bibnamefont
  {Finn}}, \bibinfo {author} {\bibfnamefont {J.-L.}\ \bibnamefont
  {Thiffeault}}, \ and\ \bibinfo {author} {\bibfnamefont {E.}~\bibnamefont
  {Gouillart}},\ }\bibfield  {title} {\enquote {\bibinfo {title} {Topological
  chaos in spatially periodic mixers},}\ }\href@noop {} {\bibfield  {journal}
  {\bibinfo  {journal} {Physica D: Nonlinear Phenomena}\ }\textbf {\bibinfo
  {volume} {221}},\ \bibinfo {pages} {92 -- 100} (\bibinfo {year}
  {2006})}\BibitemShut {NoStop}%
\bibitem [{\citenamefont {Finn}\ and\ \citenamefont
  {Thiffeault}(2011)}]{Finn:2010wg}%
  \BibitemOpen
  \bibfield  {author} {\bibinfo {author} {\bibfnamefont {M.~D.}\ \bibnamefont
  {Finn}}\ and\ \bibinfo {author} {\bibfnamefont {J.-L.}\ \bibnamefont
  {Thiffeault}},\ }\bibfield  {title} {\enquote {\bibinfo {title} {Topological
  optimisation of rod-stirring ddevices},}\ }\href@noop {} {\bibfield
  {journal} {\bibinfo  {journal} {SIAM Rev.}\ }\textbf {\bibinfo {volume}
  {53}},\ \bibinfo {pages} {723--743} (\bibinfo {year} {2011})}\BibitemShut
  {NoStop}%
\bibitem [{\citenamefont {Gouillart}, \citenamefont {Thiffeault},\ and\
  \citenamefont {Finn}(2006)}]{Gouillart:2006hi}%
  \BibitemOpen
  \bibfield  {author} {\bibinfo {author} {\bibfnamefont {E.}~\bibnamefont
  {Gouillart}}, \bibinfo {author} {\bibfnamefont {J.}~\bibnamefont
  {Thiffeault}}, \ and\ \bibinfo {author} {\bibfnamefont {M.}~\bibnamefont
  {Finn}},\ }\bibfield  {title} {\enquote {\bibinfo {title} {Topological mixing
  with ghost rods},}\ }\href@noop {} {\bibfield  {journal} {\bibinfo  {journal}
  {Phys. Rev. E}\ }\textbf {\bibinfo {volume} {73}},\ \bibinfo {pages} {036311}
  (\bibinfo {year} {2006})}\BibitemShut {NoStop}%
\bibitem [{\citenamefont {Stremler}\ \emph {et~al.}(2011)\citenamefont
  {Stremler}, \citenamefont {Ross}, \citenamefont {Grover},\ and\ \citenamefont
  {Kumar}}]{Stremler:2011hu}%
  \BibitemOpen
  \bibfield  {author} {\bibinfo {author} {\bibfnamefont {M.}~\bibnamefont
  {Stremler}}, \bibinfo {author} {\bibfnamefont {S.}~\bibnamefont {Ross}},
  \bibinfo {author} {\bibfnamefont {P.}~\bibnamefont {Grover}}, \ and\ \bibinfo
  {author} {\bibfnamefont {P.}~\bibnamefont {Kumar}},\ }\bibfield  {title}
  {\enquote {\bibinfo {title} {{Topological Chaos and Periodic Braiding of
  Almost-Cyclic Sets}},}\ }\href@noop {} {\bibfield  {journal} {\bibinfo
  {journal} {Phys. Rev. Lett.}\ }\textbf {\bibinfo {volume} {106}},\ \bibinfo
  {pages} {114101} (\bibinfo {year} {2011})}\BibitemShut {NoStop}%
\bibitem [{\citenamefont {Boyland}, \citenamefont {Stremler},\ and\
  \citenamefont {Aref}(2003)}]{Boyland:2003ux}%
  \BibitemOpen
  \bibfield  {author} {\bibinfo {author} {\bibfnamefont {P.}~\bibnamefont
  {Boyland}}, \bibinfo {author} {\bibfnamefont {M.}~\bibnamefont {Stremler}}, \
  and\ \bibinfo {author} {\bibfnamefont {H.}~\bibnamefont {Aref}},\ }\bibfield
  {title} {\enquote {\bibinfo {title} {{Topological fluid mechanics of point
  vortex motions}},}\ }\href@noop {} {\bibfield  {journal} {\bibinfo  {journal}
  {Physica D: Nonlinear Phenomena}\ }\textbf {\bibinfo {volume} {175}},\
  \bibinfo {pages} {69--95} (\bibinfo {year} {2003})}\BibitemShut {NoStop}%
\bibitem [{\citenamefont {Grover}\ \emph {et~al.}(2012)\citenamefont {Grover},
  \citenamefont {Ross}, \citenamefont {Stremler},\ and\ \citenamefont
  {Kumar}}]{Stremler2012}%
  \BibitemOpen
  \bibfield  {author} {\bibinfo {author} {\bibfnamefont {P.}~\bibnamefont
  {Grover}}, \bibinfo {author} {\bibfnamefont {S.~D.}\ \bibnamefont {Ross}},
  \bibinfo {author} {\bibfnamefont {M.~A.}\ \bibnamefont {Stremler}}, \ and\
  \bibinfo {author} {\bibfnamefont {P.}~\bibnamefont {Kumar}},\ }\bibfield
  {title} {\enquote {\bibinfo {title} {Topological chaos, braiding and
  bifurcation of almost-cyclic sets.}}\ }\href@noop {} {\bibfield  {journal}
  {\bibinfo  {journal} {Chaos}\ }\textbf {\bibinfo {volume} {22}},\ \bibinfo
  {pages} {043135 -- 043135--16} (\bibinfo {year} {2012})}\BibitemShut
  {NoStop}%
\bibitem [{\citenamefont {Vikhansky}(2004)}]{Vikhansky:2004}%
  \BibitemOpen
  \bibfield  {author} {\bibinfo {author} {\bibfnamefont {A.}~\bibnamefont
  {Vikhansky}},\ }\bibfield  {title} {\enquote {\bibinfo {title} {{Simulation
  of topological chaos in laminar flows}},}\ }\href@noop {} {\bibfield
  {journal} {\bibinfo  {journal} {AIP Chaos}\ }\textbf {\bibinfo {volume}
  {14}},\ \bibinfo {pages} {14--21} (\bibinfo {year} {2004})}\BibitemShut
  {NoStop}%
\bibitem [{\citenamefont {Finn}, \citenamefont {Cox},\ and\ \citenamefont
  {Byrne}(2003)}]{Finn:2003hd}%
  \BibitemOpen
  \bibfield  {author} {\bibinfo {author} {\bibfnamefont {M.~D.}\ \bibnamefont
  {Finn}}, \bibinfo {author} {\bibfnamefont {S.~M.}\ \bibnamefont {Cox}}, \
  and\ \bibinfo {author} {\bibfnamefont {H.~M.}\ \bibnamefont {Byrne}},\
  }\bibfield  {title} {\enquote {\bibinfo {title} {Topological chaos in
  inviscid and viscous mixers},}\ }\href@noop {} {\bibfield  {journal}
  {\bibinfo  {journal} {J. Fluid Mech.}\ }\textbf {\bibinfo {volume} {493}},\
  \bibinfo {pages} {345--361} (\bibinfo {year} {2003})}\BibitemShut {NoStop}%
\bibitem [{\citenamefont {Birman}(1975)}]{bib:BirmanBLMCG}%
  \BibitemOpen
  \bibfield  {author} {\bibinfo {author} {\bibfnamefont {J.~S.}\ \bibnamefont
  {Birman}},\ }\href@noop {} {\emph {\bibinfo {title} {Braids, Links, and
  Mapping Class Groups}}}\ (\bibinfo  {publisher} {Princeton University
  Press},\ \bibinfo {year} {1975})\BibitemShut {NoStop}%
\bibitem [{\citenamefont {Handel}(1985)}]{Handel:1985}%
  \BibitemOpen
  \bibfield  {author} {\bibinfo {author} {\bibfnamefont {M.}~\bibnamefont
  {Handel}},\ }\bibfield  {title} {\enquote {\bibinfo {title} {Gobal shadowing
  of pseudo-anosov homeomorphisms},}\ }\href@noop {} {\bibfield  {journal}
  {\bibinfo  {journal} {Ergodic Th. Dyn. Sys}\ }\textbf {\bibinfo {volume}
  {5}},\ \bibinfo {pages} {373--377} (\bibinfo {year} {1985})}\BibitemShut
  {NoStop}%
\bibitem [{\citenamefont {Thurston}(1988)}]{THURSTON:1988ty}%
  \BibitemOpen
  \bibfield  {author} {\bibinfo {author} {\bibfnamefont {W.~P.}\ \bibnamefont
  {Thurston}},\ }\bibfield  {title} {\enquote {\bibinfo {title} {{On the
  Geometry and Dynamics of Diffeomorphisms of Surfaces}},}\ }\href@noop {}
  {\bibfield  {journal} {\bibinfo  {journal} {Bulletin Of The American
  Mathematical Society}\ }\textbf {\bibinfo {volume} {19}},\ \bibinfo {pages}
  {417--431} (\bibinfo {year} {1988})}\BibitemShut {NoStop}%
\bibitem [{\citenamefont {Short}\ and\ \citenamefont
  {Wiest}(2000)}]{Short:2000wi}%
  \BibitemOpen
  \bibfield  {author} {\bibinfo {author} {\bibfnamefont {H.}~\bibnamefont
  {Short}}\ and\ \bibinfo {author} {\bibfnamefont {B.}~\bibnamefont {Wiest}},\
  }\bibfield  {title} {\enquote {\bibinfo {title} {{Orderings of Mapping Class
  Groups after Thurston}},}\ }\href@noop {} {\bibfield  {journal} {\bibinfo
  {journal} {LÕEnseignement MathŽmatique}\ }\textbf {\bibinfo {volume} {46}},\
  \bibinfo {pages} {279--312} (\bibinfo {year} {2000})}\BibitemShut {NoStop}%
\bibitem [{\citenamefont {Bernardete}, \citenamefont {Nitecki},\ and\
  \citenamefont {Gutierrez}(1995)}]{BERNARDETE:1995ua}%
  \BibitemOpen
  \bibfield  {author} {\bibinfo {author} {\bibfnamefont {D.}~\bibnamefont
  {Bernardete}}, \bibinfo {author} {\bibfnamefont {Z.}~\bibnamefont {Nitecki}},
  \ and\ \bibinfo {author} {\bibfnamefont {M.}~\bibnamefont {Gutierrez}},\
  }\bibfield  {title} {\enquote {\bibinfo {title} {{Braids and the
  Nielsen-Thurston Classification}},}\ }\href@noop {} {\bibfield  {journal}
  {\bibinfo  {journal} {Journal Of Knot Theory And Its Ramifications}\ }\textbf
  {\bibinfo {volume} {4}},\ \bibinfo {pages} {549--618} (\bibinfo {year}
  {1995})}\BibitemShut {NoStop}%
\bibitem [{\citenamefont {Bestvina}\ and\ \citenamefont
  {Handel}(1995)}]{BESTVINA:1995tx}%
  \BibitemOpen
  \bibfield  {author} {\bibinfo {author} {\bibfnamefont {M.}~\bibnamefont
  {Bestvina}}\ and\ \bibinfo {author} {\bibfnamefont {M.}~\bibnamefont
  {Handel}},\ }\bibfield  {title} {\enquote {\bibinfo {title} {{Train-Tracks
  for Surface Homeomorphisms}},}\ }\href@noop {} {\bibfield  {journal}
  {\bibinfo  {journal} {Topology}\ }\textbf {\bibinfo {volume} {34}},\ \bibinfo
  {pages} {109--140} (\bibinfo {year} {1995})}\BibitemShut {NoStop}%
\bibitem [{\citenamefont {Moussafir}(2006)}]{Moussafir}%
  \BibitemOpen
  \bibfield  {author} {\bibinfo {author} {\bibfnamefont {J.-O.}\ \bibnamefont
  {Moussafir}},\ }\bibfield  {title} {\enquote {\bibinfo {title} {On computing
  the entropy of braids},}\ }\href {\doibase 10.1007/s11853-007-0004-x}
  {\bibfield  {journal} {\bibinfo  {journal} {Functional Analysis and Other
  Mathematics}\ }\textbf {\bibinfo {volume} {1}},\ \bibinfo {pages} {37--46}
  (\bibinfo {year} {2006})}\BibitemShut {NoStop}%
\bibitem [{\citenamefont {Hall}\ and\ \citenamefont
  {Yurtta{\c{s}}}(2009)}]{Hall20091554}%
  \BibitemOpen
  \bibfield  {author} {\bibinfo {author} {\bibfnamefont {T.}~\bibnamefont
  {Hall}}\ and\ \bibinfo {author} {\bibfnamefont {S.~{\"{O}}.}\ \bibnamefont
  {Yurtta{\c{s}}}},\ }\bibfield  {title} {\enquote {\bibinfo {title} {On the
  topological entropy of families of braids},}\ }\href@noop {} {\bibfield
  {journal} {\bibinfo  {journal} {Topology and its Applications}\ }\textbf
  {\bibinfo {volume} {156}},\ \bibinfo {pages} {1554 -- 1564} (\bibinfo {year}
  {2009})}\BibitemShut {NoStop}%
\bibitem [{\citenamefont {Kassel}\ and\ \citenamefont
  {Turaev}(2008)}]{bib:Kassel}%
  \BibitemOpen
  \bibfield  {author} {\bibinfo {author} {\bibfnamefont {C.}~\bibnamefont
  {Kassel}}\ and\ \bibinfo {author} {\bibfnamefont {V.}~\bibnamefont
  {Turaev}},\ }\href@noop {} {\emph {\bibinfo {title} {Braid Groups}}}\
  (\bibinfo  {publisher} {Springer},\ \bibinfo {year} {2008})\BibitemShut
  {NoStop}%
\bibitem [{\citenamefont {Aln¾s}\ \emph {et~al.}(2015)\citenamefont {Aln¾s},
  \citenamefont {Blechta}, \citenamefont {Hake}, \citenamefont {Johansson},
  \citenamefont {Kehlet}, \citenamefont {Logg}, \citenamefont {Richardson},
  \citenamefont {Ring}, \citenamefont {Rognes},\ and\ \citenamefont
  {Wells}}]{ans20553}%
  \BibitemOpen
  \bibfield  {author} {\bibinfo {author} {\bibfnamefont {M.}~\bibnamefont
  {Aln¾s}}, \bibinfo {author} {\bibfnamefont {J.}~\bibnamefont {Blechta}},
  \bibinfo {author} {\bibfnamefont {J.}~\bibnamefont {Hake}}, \bibinfo {author}
  {\bibfnamefont {A.}~\bibnamefont {Johansson}}, \bibinfo {author}
  {\bibfnamefont {B.}~\bibnamefont {Kehlet}}, \bibinfo {author} {\bibfnamefont
  {A.}~\bibnamefont {Logg}}, \bibinfo {author} {\bibfnamefont {C.}~\bibnamefont
  {Richardson}}, \bibinfo {author} {\bibfnamefont {J.}~\bibnamefont {Ring}},
  \bibinfo {author} {\bibfnamefont {M.}~\bibnamefont {Rognes}}, \ and\ \bibinfo
  {author} {\bibfnamefont {G.}~\bibnamefont {Wells}},\ }\bibfield  {title}
  {\enquote {\bibinfo {title} {The fenics project version 1.5},}\ }\href
  {\doibase 10.11588/ans.2015.100.20553} {\bibfield  {journal} {\bibinfo
  {journal} {Archive of Numerical Software}\ }\textbf {\bibinfo {volume} {3}}
  (\bibinfo {year} {2015}),\ 10.11588/ans.2015.100.20553}\BibitemShut {NoStop}%
\bibitem [{\citenamefont {Logg}, \citenamefont {Mardal},\ and\ \citenamefont
  {Wells}(2012)}]{bib:FEniCS}%
  \BibitemOpen
  \bibinfo {editor} {\bibfnamefont {A.}~\bibnamefont {Logg}}, \bibinfo {editor}
  {\bibfnamefont {K.-A.}\ \bibnamefont {Mardal}}, \ and\ \bibinfo {editor}
  {\bibfnamefont {G.~N.}\ \bibnamefont {Wells}},\ eds.,\ \href@noop {} {\emph
  {\bibinfo {title} {Automated Solutions of Differential Equations by the
  Finite Element Method: The FEniCS Book}}}\ (\bibinfo  {publisher}
  {Springer},\ \bibinfo {year} {2012})\BibitemShut {NoStop}%
\bibitem [{\citenamefont {Lai}\ and\ \citenamefont
  {Peskin}(2000)}]{Lai2000705}%
  \BibitemOpen
  \bibfield  {author} {\bibinfo {author} {\bibfnamefont {M.-C.}\ \bibnamefont
  {Lai}}\ and\ \bibinfo {author} {\bibfnamefont {C.~S.}\ \bibnamefont
  {Peskin}},\ }\bibfield  {title} {\enquote {\bibinfo {title} {An immersed
  boundary method with formal second-order accuracy and reduced numerical
  viscosity},}\ }\href@noop {} {\bibfield  {journal} {\bibinfo  {journal}
  {Journal of Computational Physics}\ }\textbf {\bibinfo {volume} {160}},\
  \bibinfo {pages} {705 -- 719} (\bibinfo {year} {2000})}\BibitemShut {NoStop}%
\bibitem [{\citenamefont {Chorin}(1968)}]{Chorin}%
  \BibitemOpen
  \bibfield  {author} {\bibinfo {author} {\bibfnamefont {A.~J.}\ \bibnamefont
  {Chorin}},\ }\bibfield  {title} {\enquote {\bibinfo {title} {Numerical
  solution of the navier-stokes equation},}\ }\href@noop {} {\bibfield
  {journal} {\bibinfo  {journal} {Math. Comp.}\ }\textbf {\bibinfo {volume}
  {22}},\ \bibinfo {pages} {745--762} (\bibinfo {year} {1968})}\BibitemShut
  {NoStop}%
\bibitem [{\citenamefont {Ottino}, \citenamefont {Jana},\ and\ \citenamefont
  {Chakravarthy}(1994)}]{Ottino1994}%
  \BibitemOpen
  \bibfield  {author} {\bibinfo {author} {\bibfnamefont {J.~M.}\ \bibnamefont
  {Ottino}}, \bibinfo {author} {\bibfnamefont {S.~C.}\ \bibnamefont {Jana}}, \
  and\ \bibinfo {author} {\bibfnamefont {V.~S.}\ \bibnamefont {Chakravarthy}},\
  }\bibfield  {title} {\enquote {\bibinfo {title} {From reynolds's stretching
  and folding to mixing studies using horseshoe maps},}\ }\href {\doibase
  http://dx.doi.org/10.1063/1.868308} {\bibfield  {journal} {\bibinfo
  {journal} {Physics of Fluids}\ }\textbf {\bibinfo {volume} {6}},\ \bibinfo
  {pages} {685--699} (\bibinfo {year} {1994})}\BibitemShut {NoStop}%
\bibitem [{\citenamefont {Ottino}(1989)}]{ottino1989kinematics}%
  \BibitemOpen
  \bibfield  {author} {\bibinfo {author} {\bibfnamefont {J.}~\bibnamefont
  {Ottino}},\ }\href {https://books.google.com/books?id=8OLVcbRoNSgC} {\emph
  {\bibinfo {title} {The Kinematics of Mixing: Stretching, Chaos, and
  Transport}}},\ Cambridge Texts in Applied Mathematics\ (\bibinfo  {publisher}
  {Cambridge University Press},\ \bibinfo {year} {1989})\BibitemShut {NoStop}%
\bibitem [{\citenamefont {Lapeyre}(2002)}]{LapeyreFTLE2002}%
  \BibitemOpen
  \bibfield  {author} {\bibinfo {author} {\bibfnamefont {G.}~\bibnamefont
  {Lapeyre}},\ }\bibfield  {title} {\enquote {\bibinfo {title}
  {Characterization of finite-time lyapunov exponents and vectors in
  two-dimensional turbulence},}\ }\href {\doibase
  http://dx.doi.org/10.1063/1.1499395} {\bibfield  {journal} {\bibinfo
  {journal} {Chaos}\ }\textbf {\bibinfo {volume} {12}},\ \bibinfo {pages}
  {688--698} (\bibinfo {year} {2002})}\BibitemShut {NoStop}%
\bibitem [{\citenamefont {Voth}, \citenamefont {Haller},\ and\ \citenamefont
  {Gollub}(2002)}]{VothHallerGollubExperiment2002}%
  \BibitemOpen
  \bibfield  {author} {\bibinfo {author} {\bibfnamefont {G.~A.}\ \bibnamefont
  {Voth}}, \bibinfo {author} {\bibfnamefont {G.}~\bibnamefont {Haller}}, \ and\
  \bibinfo {author} {\bibfnamefont {J.~P.}\ \bibnamefont {Gollub}},\ }\bibfield
   {title} {\enquote {\bibinfo {title} {Experimental measurements of stretching
  fields in fluid mixing},}\ }\href {\doibase 10.1103/PhysRevLett.88.254501}
  {\bibfield  {journal} {\bibinfo  {journal} {Phys. Rev. Lett.}\ }\textbf
  {\bibinfo {volume} {88}},\ \bibinfo {pages} {254501} (\bibinfo {year}
  {2002})}\BibitemShut {NoStop}%
\bibitem [{\citenamefont {Tang}\ and\ \citenamefont
  {Boozer}(1996)}]{FTLEReactionDiffusionTang1996}%
  \BibitemOpen
  \bibfield  {author} {\bibinfo {author} {\bibfnamefont {X.}~\bibnamefont
  {Tang}}\ and\ \bibinfo {author} {\bibfnamefont {A.}~\bibnamefont {Boozer}},\
  }\bibfield  {title} {\enquote {\bibinfo {title} {Finite time lyapunov
  exponent and advection-diffusion equation},}\ }\href {\doibase
  http://dx.doi.org/10.1016/0167-2789(96)00064-4} {\bibfield  {journal}
  {\bibinfo  {journal} {Physica D: Nonlinear Phenomena}\ }\textbf {\bibinfo
  {volume} {95}},\ \bibinfo {pages} {283 -- 305} (\bibinfo {year}
  {1996})}\BibitemShut {NoStop}%
\bibitem [{\citenamefont {Thiffeault}\ and\ \citenamefont
  {Boozer}(2001)}]{FTLEgeometryThiffeault2001}%
  \BibitemOpen
  \bibfield  {author} {\bibinfo {author} {\bibfnamefont {J.-L.}\ \bibnamefont
  {Thiffeault}}\ and\ \bibinfo {author} {\bibfnamefont {A.~H.}\ \bibnamefont
  {Boozer}},\ }\bibfield  {title} {\enquote {\bibinfo {title} {Geometrical
  constraints on finite-time lyapunov exponents in two and three dimensions},}\
  }\href {\doibase http://dx.doi.org/10.1063/1.1342079} {\bibfield  {journal}
  {\bibinfo  {journal} {Chaos}\ }\textbf {\bibinfo {volume} {11}},\ \bibinfo
  {pages} {16--28} (\bibinfo {year} {2001})}\BibitemShut {NoStop}%
\bibitem [{\citenamefont {Yuan}\ \emph {et~al.}(2000)\citenamefont {Yuan},
  \citenamefont {Nam}, \citenamefont {Antonsen}, \citenamefont {Ott},\ and\
  \citenamefont {Guzdar}}]{YuanOtt2000}%
  \BibitemOpen
  \bibfield  {author} {\bibinfo {author} {\bibfnamefont {G.-C.}\ \bibnamefont
  {Yuan}}, \bibinfo {author} {\bibfnamefont {K.}~\bibnamefont {Nam}}, \bibinfo
  {author} {\bibfnamefont {T.~M.}\ \bibnamefont {Antonsen}}, \bibinfo {author}
  {\bibfnamefont {E.}~\bibnamefont {Ott}}, \ and\ \bibinfo {author}
  {\bibfnamefont {P.~N.}\ \bibnamefont {Guzdar}},\ }\bibfield  {title}
  {\enquote {\bibinfo {title} {Power spectrum of passive scalars in two
  dimensional chaotic flows},}\ }\href {\doibase
  http://dx.doi.org/10.1063/1.166474} {\bibfield  {journal} {\bibinfo
  {journal} {Chaos}\ }\textbf {\bibinfo {volume} {10}},\ \bibinfo {pages}
  {39--49} (\bibinfo {year} {2000})}\BibitemShut {NoStop}%
\bibitem [{\citenamefont {Antonsen}\ \emph {et~al.}(1996)\citenamefont
  {Antonsen}, \citenamefont {Fan}, \citenamefont {Ott},\ and\ \citenamefont
  {Garcia‐Lopez}}]{ChaoticOrbitsPowerSpectrum}%
  \BibitemOpen
  \bibfield  {author} {\bibinfo {author} {\bibfnamefont {T.~M.}\ \bibnamefont
  {Antonsen}}, \bibinfo {author} {\bibfnamefont {Z.}~\bibnamefont {Fan}},
  \bibinfo {author} {\bibfnamefont {E.}~\bibnamefont {Ott}}, \ and\ \bibinfo
  {author} {\bibfnamefont {E.}~\bibnamefont {Garcia‐Lopez}},\ }\bibfield
  {title} {\enquote {\bibinfo {title} {The role of chaotic orbits in the
  determination of power spectra of passive scalars},}\ }\href {\doibase
  http://dx.doi.org/10.1063/1.869083} {\bibfield  {journal} {\bibinfo
  {journal} {Physics of Fluids}\ }\textbf {\bibinfo {volume} {8}},\ \bibinfo
  {pages} {3094--3104} (\bibinfo {year} {1996})}\BibitemShut {NoStop}%
\bibitem [{\citenamefont {Sarkar}, \citenamefont {Narv{\'a}ez},\ and\
  \citenamefont {Harting}(2012)}]{FTLEmicromixersSarkar2012}%
  \BibitemOpen
  \bibfield  {author} {\bibinfo {author} {\bibfnamefont {A.}~\bibnamefont
  {Sarkar}}, \bibinfo {author} {\bibfnamefont {A.}~\bibnamefont {Narv{\'a}ez}},
  \ and\ \bibinfo {author} {\bibfnamefont {J.}~\bibnamefont {Harting}},\
  }\bibfield  {title} {\enquote {\bibinfo {title} {Quantification of the
  performance of chaotic micromixers on the basis of finite time lyapunov
  exponents},}\ }\href {\doibase 10.1007/s10404-012-0936-4} {\bibfield
  {journal} {\bibinfo  {journal} {Microfluidics and Nanofluidics}\ }\textbf
  {\bibinfo {volume} {13}},\ \bibinfo {pages} {19--27} (\bibinfo {year}
  {2012})}\BibitemShut {NoStop}%
\bibitem [{\citenamefont {Bowman}(1993)}]{GeophysicsFTLEBowman1993}%
  \BibitemOpen
  \bibfield  {author} {\bibinfo {author} {\bibfnamefont {K.~P.}\ \bibnamefont
  {Bowman}},\ }\bibfield  {title} {\enquote {\bibinfo {title} {Large-scale
  isentropic mixing properties of the antarctic polar vortex from analyzed
  winds},}\ }\href {\doibase 10.1029/93JD02599} {\bibfield  {journal} {\bibinfo
   {journal} {Journal of Geophysical Research: Atmospheres}\ }\textbf {\bibinfo
  {volume} {98}},\ \bibinfo {pages} {23013--23027} (\bibinfo {year}
  {1993})}\BibitemShut {NoStop}%
\bibitem [{\citenamefont {Arratia}\ and\ \citenamefont
  {Gollub}(2006)}]{GollubPhysRevLett2006}%
  \BibitemOpen
  \bibfield  {author} {\bibinfo {author} {\bibfnamefont {P.~E.}\ \bibnamefont
  {Arratia}}\ and\ \bibinfo {author} {\bibfnamefont {J.~P.}\ \bibnamefont
  {Gollub}},\ }\bibfield  {title} {\enquote {\bibinfo {title} {Predicting the
  progress of diffusively limited chemical reactions in the presence of chaotic
  advection},}\ }\href {\doibase 10.1103/PhysRevLett.96.024501} {\bibfield
  {journal} {\bibinfo  {journal} {Phys. Rev. Lett.}\ }\textbf {\bibinfo
  {volume} {96}},\ \bibinfo {pages} {024501} (\bibinfo {year}
  {2006})}\BibitemShut {NoStop}%
\bibitem [{\citenamefont {Cvitanovi\'c}\ \emph {et~al.}(2012)\citenamefont
  {Cvitanovi\'c}, \citenamefont {Artuso}, \citenamefont {Mainieri},
  \citenamefont {Tanner},\ and\ \citenamefont {Vattay}}]{ChaosBook}%
  \BibitemOpen
  \bibfield  {author} {\bibinfo {author} {\bibfnamefont {P.}~\bibnamefont
  {Cvitanovi\'c}}, \bibinfo {author} {\bibfnamefont {R.}~\bibnamefont
  {Artuso}}, \bibinfo {author} {\bibfnamefont {R.}~\bibnamefont {Mainieri}},
  \bibinfo {author} {\bibfnamefont {G.}~\bibnamefont {Tanner}}, \ and\ \bibinfo
  {author} {\bibfnamefont {G.}~\bibnamefont {Vattay}},\ }\href@noop {} {\emph
  {\bibinfo {title} {Chaos: Classical and Quantum}}}\ (\bibinfo  {publisher}
  {Niels Bohr Institute},\ \bibinfo {address} {Copenhagen},\ \bibinfo {year}
  {2012})\ \bibinfo {note} {{\tt
  \href{http://ChaosBook.org}{ChaosBook.org}}}\BibitemShut {NoStop}%
\bibitem [{\citenamefont {Benettin}, \citenamefont {Galgani},\ and\
  \citenamefont {Strelcyn}(1976)}]{Benettin76}%
  \BibitemOpen
  \bibfield  {author} {\bibinfo {author} {\bibfnamefont {G.}~\bibnamefont
  {Benettin}}, \bibinfo {author} {\bibfnamefont {L.}~\bibnamefont {Galgani}}, \
  and\ \bibinfo {author} {\bibfnamefont {J.-M.}\ \bibnamefont {Strelcyn}},\
  }\bibfield  {title} {\enquote {\bibinfo {title} {Kolmogorov entropy and
  numerical experiments},}\ }\href {\doibase 10.1103/PhysRevA.14.2338}
  {\bibfield  {journal} {\bibinfo  {journal} {Phys. Rev. A}\ }\textbf {\bibinfo
  {volume} {14}},\ \bibinfo {pages} {2338--2345} (\bibinfo {year}
  {1976})}\BibitemShut {NoStop}%
\bibitem [{\citenamefont {Shimada}\ and\ \citenamefont
  {Nagashima}(1979)}]{Shimada79}%
  \BibitemOpen
  \bibfield  {author} {\bibinfo {author} {\bibfnamefont {I.}~\bibnamefont
  {Shimada}}\ and\ \bibinfo {author} {\bibfnamefont {T.}~\bibnamefont
  {Nagashima}},\ }\bibfield  {title} {\enquote {\bibinfo {title} {A numerical
  approach to ergodic problem of dissipative dynamical systems},}\ }\href
  {\doibase 10.1143/PTP.61.1605} {\bibfield  {journal} {\bibinfo  {journal}
  {Progress of Theoretical Physics}\ }\textbf {\bibinfo {volume} {61}},\
  \bibinfo {pages} {1605--1616} (\bibinfo {year} {1979})}\BibitemShut {NoStop}%
\bibitem [{\citenamefont {Xia}\ \emph {et~al.}(2006)\citenamefont {Xia},
  \citenamefont {Shu}, \citenamefont {Wan},\ and\ \citenamefont
  {Chew}}]{micromixerReynoldsXia2006}%
  \BibitemOpen
  \bibfield  {author} {\bibinfo {author} {\bibfnamefont {H.~M.}\ \bibnamefont
  {Xia}}, \bibinfo {author} {\bibfnamefont {C.}~\bibnamefont {Shu}}, \bibinfo
  {author} {\bibfnamefont {S.~Y.~M.}\ \bibnamefont {Wan}}, \ and\ \bibinfo
  {author} {\bibfnamefont {Y.~T.}\ \bibnamefont {Chew}},\ }\bibfield  {title}
  {\enquote {\bibinfo {title} {Influence of the reynolds number on chaotic
  mixing in a spatially periodic micromixer and its characterization using
  dynamical system techniques},}\ }\href
  {http://stacks.iop.org/0960-1317/16/i=1/a=008} {\bibfield  {journal}
  {\bibinfo  {journal} {Journal of Micromechanics and Microengineering}\
  }\textbf {\bibinfo {volume} {16}},\ \bibinfo {pages} {53} (\bibinfo {year}
  {2006})}\BibitemShut {NoStop}%
\bibitem [{\citenamefont {Wang}\ \emph {et~al.}(2009)\citenamefont {Wang},
  \citenamefont {Feng}, \citenamefont {Ottino},\ and\ \citenamefont
  {Lueptow}}]{ChaoticAdvecInertiaWang2009}%
  \BibitemOpen
  \bibfield  {author} {\bibinfo {author} {\bibfnamefont {J.}~\bibnamefont
  {Wang}}, \bibinfo {author} {\bibfnamefont {L.}~\bibnamefont {Feng}}, \bibinfo
  {author} {\bibfnamefont {J.~M.}\ \bibnamefont {Ottino}}, \ and\ \bibinfo
  {author} {\bibfnamefont {R.}~\bibnamefont {Lueptow}},\ }\bibfield  {title}
  {\enquote {\bibinfo {title} {Inertial effects on chaotic advection and mixing
  in a 2d cavity flow},}\ }\href {\doibase 10.1021/ie800404d} {\bibfield
  {journal} {\bibinfo  {journal} {Industrial \& Engineering Chemistry
  Research}\ }\textbf {\bibinfo {volume} {48}},\ \bibinfo {pages} {2436--2442}
  (\bibinfo {year} {2009})},\ \Eprint
  {http://arxiv.org/abs/http://dx.doi.org/10.1021/ie800404d}
  {http://dx.doi.org/10.1021/ie800404d} \BibitemShut {NoStop}%
\bibitem [{\citenamefont {Hobbs}\ and\ \citenamefont
  {Muzzio}(1998)}]{KinicsMixerReDepHobbs1998}%
  \BibitemOpen
  \bibfield  {author} {\bibinfo {author} {\bibfnamefont {D.}~\bibnamefont
  {Hobbs}}\ and\ \bibinfo {author} {\bibfnamefont {F.}~\bibnamefont {Muzzio}},\
  }\bibfield  {title} {\enquote {\bibinfo {title} {Reynolds number effects on
  laminar mixing in the kenics static mixer},}\ }\href {\doibase
  http://dx.doi.org/10.1016/S0923-0467(98)00065-7} {\bibfield  {journal}
  {\bibinfo  {journal} {Chemical Engineering Journal}\ }\textbf {\bibinfo
  {volume} {70}},\ \bibinfo {pages} {93 -- 104} (\bibinfo {year}
  {1998})}\BibitemShut {NoStop}%
\bibitem [{\citenamefont {Clifford}, \citenamefont {Cox},\ and\ \citenamefont
  {Finn}(2004)}]{PanetaryMixerClifford2004}%
  \BibitemOpen
  \bibfield  {author} {\bibinfo {author} {\bibfnamefont {M.}~\bibnamefont
  {Clifford}}, \bibinfo {author} {\bibfnamefont {S.}~\bibnamefont {Cox}}, \
  and\ \bibinfo {author} {\bibfnamefont {M.}~\bibnamefont {Finn}},\ }\bibfield
  {title} {\enquote {\bibinfo {title} {Reynolds number effects in a simple
  planetary mixer},}\ }\href {\doibase
  http://dx.doi.org/10.1016/j.ces.2004.03.043} {\bibfield  {journal} {\bibinfo
  {journal} {Chemical Engineering Science}\ }\textbf {\bibinfo {volume} {59}},\
  \bibinfo {pages} {3371 -- 3379} (\bibinfo {year} {2004})}\BibitemShut
  {NoStop}%
\bibitem [{\citenamefont {{Mezic}}(2001)}]{2001Mezic}%
  \BibitemOpen
  \bibfield  {author} {\bibinfo {author} {\bibfnamefont {I.}~\bibnamefont
  {{Mezic}}},\ }\bibfield  {title} {\enquote {\bibinfo {title} {{Chaotic
  advection in bounded Navier Stokes flows}},}\ }\href@noop {} {\bibfield
  {journal} {\bibinfo  {journal} {Journal of Fluid Mechanics}\ }\textbf
  {\bibinfo {volume} {431}},\ \bibinfo {pages} {347--370} (\bibinfo {year}
  {2001})}\BibitemShut {NoStop}%
\bibitem [{\citenamefont {Baggaley}, \citenamefont {Barenghi},\ and\
  \citenamefont {Shukurov}(2009)}]{LEvReTurbBaggaley2009}%
  \BibitemOpen
  \bibfield  {author} {\bibinfo {author} {\bibfnamefont {A.~W.}\ \bibnamefont
  {Baggaley}}, \bibinfo {author} {\bibfnamefont {C.~F.}\ \bibnamefont
  {Barenghi}}, \ and\ \bibinfo {author} {\bibfnamefont {A.}~\bibnamefont
  {Shukurov}},\ }\bibfield  {title} {\enquote {\bibinfo {title} {Stretching in
  a model of a turbulent flow},}\ }\href {\doibase
  http://dx.doi.org/10.1016/j.physd.2008.10.013} {\bibfield  {journal}
  {\bibinfo  {journal} {Physica D: Nonlinear Phenomena}\ }\textbf {\bibinfo
  {volume} {238}},\ \bibinfo {pages} {365 -- 369} (\bibinfo {year}
  {2009})}\BibitemShut {NoStop}%
\end{thebibliography}%

\end{document}